\documentclass[format=acmsmall, review=false, screen=true]{acmart}

\usepackage{booktabs} % For formal tables
\usepackage{tikz}
\usepackage{algorithm2e}
\usepackage{multirow}
\usepackage{csquotes}
\usepackage{tabularx}
\usepackage{subfigure}
\usepackage{graphicx}
\usepackage{caption}
\usepackage{xspace}
\usepackage{listings}
\usepackage{colortbl}
\usepackage{microtype}
\usepackage[inline,shortlabels]{enumitem}
\usepackage{color}
\usepackage{enumerate}
\usepackage{wrapfig,lipsum,booktabs}

\SetAlFnt{\small}
\SetAlCapFnt{\small}
\SetAlCapNameFnt{\small}
\SetAlCapHSkip{0pt}
\IncMargin{-\parindent}

\acmJournal{JETC}
\acmMonth{9}
\copyrightyear{2019}
\begin{document}
% Title portion. Note the short title for running heads
\title[QuTiBench: Benchmarking Neural Networks on Heterogeneous Hardware]{QuTiBench: Benchmarking  Neural Networks on Heterogeneous Hardware}

\author{Michaela Blott}
\email{michaela.blott@xilinx.com}
\affiliation{
  \institution{Xilinx Research}
  \city{Dublin}
  \country{Ireland}
}
\author{Lisa Halder}
\email{lisa.halder@xilinx.com}
\affiliation{
  \institution{Xilinx Research, Ulm University}
  \city{Dublin}
  \country{Ireland}
}
\author{Miriam Leeser}
\email{mel@coe.neu.edu}
\affiliation{
  \institution{Northeastern University}
  \city{Boston}
  \state{Massachusetts}
  \country{US}
}
\author{Linda Doyle}
\email{}
\affiliation{
\institution{Trinity College Dublin}
\city{Dublin}
\country{Ireland}
}

\begin{abstract}
Neural Networks have become one of the most successful
universal machine learning algorithms. They play a key role in enabling machine vision and speech recognition, and are increasingly adopted in other application domains. Their computational complexity is enormous and comes along with
equally challenging memory requirements both in regards to capacity and access bandwidth, which limits deployment in particular within energy constrained, embedded environments.  
In order to address these implementation challenges, a broad spectrum of new customized and heterogeneous hardware architectures have emerged, 
often accompanied with co-designed algorithms to extract maximum benefit out of the hardware. Furthermore, numerous optimization techniques are being explored for neural networks to reduce compute and memory requirements while maintaining accuracy.
This results in an abundance of algorithmic and architectural choices, some of which fit specific use cases better than others.  

For system level designers, there is currently no good way to compare the variety of hardware, algorithm and optimization options. While there are many benchmarking efforts in this field, they cover only subsections of the embedded design space.  None of the existing benchmarks support essential algorithmic optimizations such as quantization, an important technique to stay on chip, or specialized heterogeneous hardware architectures. We propose a novel benchmark suite, {\em QuTiBench}, that addresses this need. QuTiBench is a novel multi-tiered benchmarking methodology (\emph{Ti}) that supports algorithmic optimizations such as quantization (\emph{Qu}) and helps system developers understand the benefits and limitations of these novel compute architectures in regard to specific neural networks and will help drive future innovation.  We invite the community to contribute to QuTiBench in order to support the full spectrum of choices in implementing machine learning systems.

\end{abstract}



\begin{comment}
\begin{CCSXML}
<ccs2012>
 <concept>
  <concept_id>10010520.10010553.10010562</concept_id>
  <concept_desc>Computer systems organization~Embedded systems</concept_desc>
  <concept_significance>500</concept_significance>
 </concept>
 <concept>
  <concept_id>10010520.10010575.10010755</concept_id>
  <concept_desc>Computer systems organization~Redundancy</concept_desc>
  <concept_significance>300</concept_significance>
 </concept>
 <concept>
  <concept_id>10010520.10010553.10010554</concept_id>
  <concept_desc>Computer systems organization~Robotics</concept_desc>
  <concept_significance>100</concept_significance>
 </concept>
 <concept>
  <concept_id>10003033.10003083.10003095</concept_id>
  <concept_desc>Networks~Network reliability</concept_desc>
  <concept_significance>100</concept_significance>
 </concept>
</ccs2012>
\end{CCSXML}

\ccsdesc[500]{Computer systems organization~Embedded systems}
\ccsdesc[300]{Computer systems organization~Redundancy}
\ccsdesc{Computer systems organization~Robotics}
\ccsdesc[100]{Networks~Network reliability}

\end{comment}


\maketitle


\renewcommand{\shortauthors}{M. Blott et al.}

\section{Introduction}
\label{sec:introduction}

Over the last several years, neural networks (NNs)\footnote{We use the terms neural network and model synonymously throughout this article} have become incredibly successful.
A huge variety of neural networks are increasingly deployed in conjunction with robotics, advanced driver assistance systems (ADAS), security monitors and many other applications. 
Furthermore, as they have the theoretical property of being a universal approximator which requires zero domain expertise, they are increasingly applied to previously unsolved problems, and sometimes to replace existing algorithms, unless of course the original algorithm is of much lower complexity.  Note that the applications listed above are all embedded applications, and there is an increasing interest in training as well as inference in such environments.  

\begin{wrapfigure}{r}{0.6\linewidth}
  \caption{Accuracy-Hardware Cost Tradeoffs}
  \begin{center}
   \includegraphics[width=0.55\textwidth]{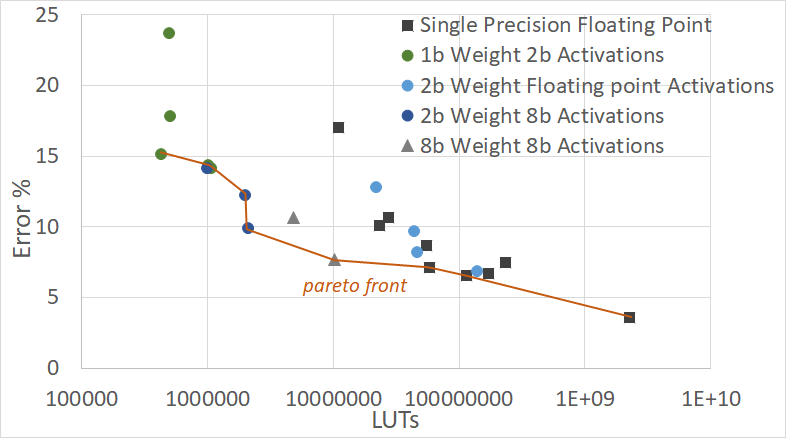}
  \end{center}
  \label{fig:pareto1}
\end{wrapfigure}

The challenge of deploying these networks lies in their compute and memory intensity, which poses the largest barrier to adoption particularly within the embedded space where compute resources, power and memory are at premium.
%According to researchers at Baidu, 
Inference requires often billions of operations and training for modern algorithms involves tens of single-precision exaflops to converge and has tens of millions of parameters \cite{amodei2016deep}.  
The interest to apply these techniques in energy constrained environments has spawned a rise in algorithmic and architectural innovation.
Algorithmic optimizations include topological transformations with pruning and compression schemes. In addition, the general trend towards transprecision computing \cite{tagliavini2018transprecision,malossi2018transprecision} can be nicely exploited within this particular application context. Extreme reduced precision neural networks for example, which take datatypes down to ternary or even binary representations can bring significant hardware cost savings and minimal accuracy impact, as visualized in Fig.~\ref{fig:pareto1}\cite{iccd}.

Architectural innovation is showcased by Google's TPU \cite{tpu}, numerous start-up companies such as Nervana, Graphcore, GROC, and Cerebras, as well as a spectrum of reconfigurable accelerators leveraging FPGAs. 
Each of these architectures brings their own inherent benefit. Overall, it is becoming increasingly difficult to predict which architecture will deliver what performance for which particular neural network. This poses the key challenge that we address with our benchmark suite. 

Benchmarks at their core encompass a suite of tests for evaluating performance or level of quality.
When done well, benchmarking creates clarity
by establishing fair baselines and providing representative comparisons between different platforms and compute fabrics. 
They act as the antidote to product marketing and provide system designers a toolbox to avoid making poor choices where end systems fail to meet requirements such as throughput, power or cost, and delay product launch.
The benefits of a good benchmarking suite go beyond this and provide insights from all perspectives. 
Benchmarks can be of high benefit to hardware designers as well as end users. Benchmarks drive optimizations for semiconductor companies who are customizing compute fabrics for deep learning applications, and for end users standardized tests help drive optimal purchasing choices. Finally, for newcomers to the domain, benchmarking suites can offer objective summaries that introduce key figures of merit and basic choices as well as setting expectations of the state of the art. 

\begin{figure}
\centering
\includegraphics[width=0.9\linewidth]{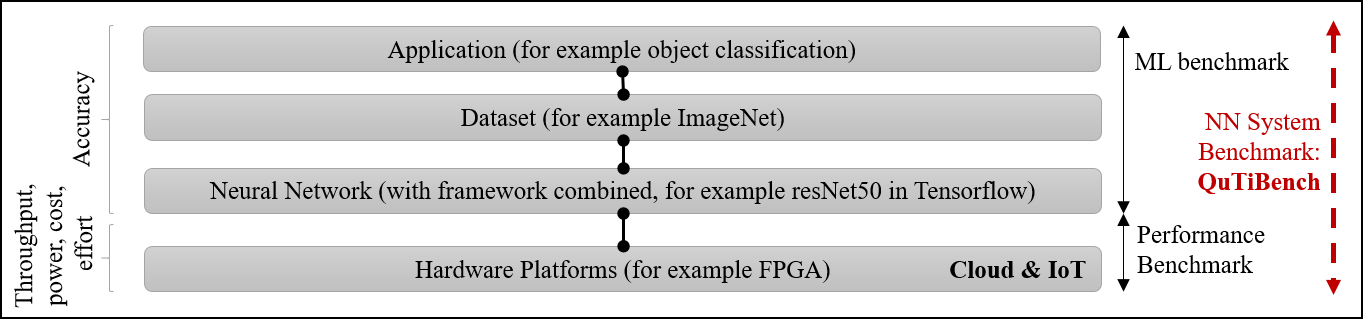}
\caption{Multidimensional Design Space}
\label{fig:intro}
  \vspace*{-3ex}
\end{figure}

This is an extremely complex design space to visualize, as shown in Fig.~\ref{fig:intro}.
There are numerous \emph{machine learning applications}, and each of these can be trained with different \emph{datasets} and different neural network \emph{models} and variations, and depending on these factors (as well as numerical representations, learning techniques and hyperparameter selection) can produce different results, the key figure of merit being test error rate or conversely, \emph{accuracy}. 
There are numerous choices with different \emph{hardware platforms} within the cloud and IoT spaces and everywhere in between. 
All of the implementation alternatives will deliver different \emph{performance} in tera or giga operations per second (TOP/s or GOP/s), \emph{response time}, \emph{power consumption}, \emph{cost} and required \emph{development effort}.

Within this space there are two main types of benchmarks: \emph{Machine Learning (ML) benchmarks} and \emph{performance benchmarks}.
ML benchmarks are typically aimed at achieving low test error, independent of the hardware implications, therefore being of limited efficiency.
Examples are the ILSVRC ImageNet competition, as well as more sophisticated efforts such as MLBench~\cite{mlbench}.
Performance benchmarks are agnostic of the target application, measuring performance characteristics such as throughput and power for characteristic compute patterns. Even when tailored towards characteristic ML workloads, they do not capture the fact that for different hardware architectures, different compute patterns should be used. Most importantly, they do not correlate their results regarding algorithmic optimization  back to the application level target, which is accuracy, and therefore provide the necessary freedom and scope for algorithmic modifications, an essential ingredient to extracting performance out of heterogeneous computing systems.

In this paper we present \emph{QuTiBench}, a benchmarking suite that lies at the intersection of the machine learning and  hardware communities and spans the full design space. QuTiBench couples neural network performance with hardware performance and as such can provide insights as to what is the best possible combination within this design space for specific use cases.
Although there are a number of efforts emerging in this space, such as DeepBench and MLPerf, there is currently no comprehensive benchmarking suite in existence that addresses the scope of what is needed, and in particular targets embedded systems. QuTiBench is unique in the way we support quantization (\emph{Qu}) which is an important optimization technique for neural networks and leveraged by many specialized hardware architectures. Furthermore, QuTiBench provides multiple tiers of tests (\emph{Ti}) which can provide deep insights for the composition of complex systems and provide tradeoffs between speed and accuracy across a broad range of systems.  

The main contribution of this paper is the definition of QuTiBench, which has the following unique features:

\noindent $\bullet$ It is a multi-tiered approach that supports a range of compromises for benchmarking in regards to quality of prediction and effort.  In particular, QuTiBench supports theoretical results as a measuring stick, different computational patterns for different neural networks, and combinations of microbenchmarks and full applications for addressing the end user design space. 

\noindent $\bullet$It supports algorithmic optimizations and levels of development effort including naive and optimized implementations, by correlating everything at the application level's figures of merit.
%, and visualizes the full design space. 

\noindent $\bullet$ In particular, QuTiBench supports different approaches to quantization at all levels, which is essential for efficient, low power architectures.

\noindent $\bullet$ It supports a broad range of applications, both inference and training, and available systems from cloud to IoT. 
    
QuTiBench is still in its early stages.  We hope the community will help make this a valuable contribution to the Machine Learning field.  In this paper we provide the first analysis of theoretical compute and memory requirements for both applications and candidate hardware platforms, which forms level 0 of our benchmark suite. We present initial experimental results to validate the benchmarking methodology, as well as outline plans for the remaining levels.  

The remainder of this article is structured as follows: We start with background on neural networks.  Sec.~\ref{sec:cmpmem} analyses the compute and memory requirements of a broad selection of networks. Sec.~\ref{sec:hw} provides details on different hardware architectures and how inference and training workloads can be mapped to them. This provides insights into the spectrum of implementation choices and how they are represented within the benchmark suite.
In Sec.~\ref{sec:whybench} we take a closer look at the key components, characteristics and challenges of a benchmarking suite in Machine Learning. Sec.~\ref{sec:rel} describes existing efforts in this space and  Sec.~\ref{sec:proposal} introduces the key concepts of QuTiBench.  We evaluate our approach with experimental results in Sec.~\ref{sec:experiment}. Sec.~\ref{sec:Con} concludes the article and presents future directions. Full experimental results can be found in the appendix.

 %Michaela
\section{Background on Neural Networks}\label{secNN}

This effort focuses on \emph{neural networks (NNs)}, a class of machine learning algorithms that forms a subclass of artificial intelligence. With its property of being a universal approximator \cite{approximation}, NNs increasingly outperform and replace existing algorithms. 
NNs can also provide automation for previously unsolved applications, where no algorithms exist.
No domain expertise is required, just sufficiently large datasets together with a sufficiently large topology for the network to train for a given accuracy target. These factors contribute to NN's popularity.

\begin{figure}
\centering
\includegraphics[width=0.7\linewidth]{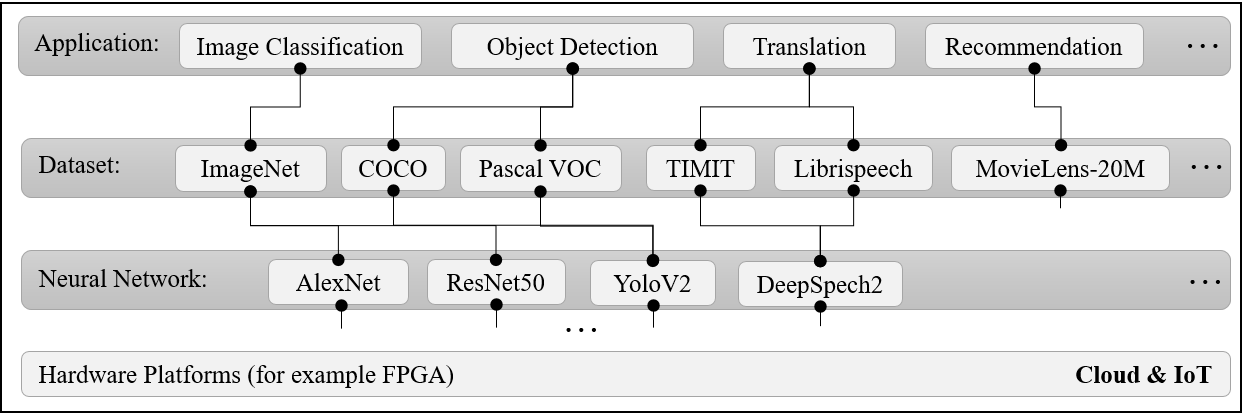}
\caption{Applications, Datasets, Neural Networks}
\label{fig:fig3apps}
  \vspace*{-2ex}
\end{figure}

The design space (see Fig.~\ref{fig:fig3apps}) is complex. For every application there are many different types of NNs, and new algorithms continue to evolve. Furthermore, different types of datasets can be used. The resulting combinations can achieve different accuracy targets, and are accompanied by different compute requirements. Also, a neural network model is always paired with the particular framework in which it was trained, which can have impact on the accuracy.  

There is a large application space for neural networks (see Table~\ref{table:nnoverview}) with domains ranging from vision to natural language processing (NLP) to gaming and recommendation systems. In each domain, there are numerous tasks which are amenable for neural networks; for example, within the vision processing context: image classification, object detection, and semantic segmentation. Furthermore, these models can be trained using different training techniques. 
Note that it is not easy to define clear categories as terms overlap. For example, deep reinforcement learning techniques can be applied to any network. Seq2Seq networks is a full family of networks, while ResNet50, VGG, and InceptionV3 refer to specific topologies.

\begin{table}
\caption{Breadth of popular ML Tasks and NN Types}
\label{table:nnoverview}
  \resizebox{\linewidth}{!}{
\begin{tabular}{|lll|cc|}
\toprule
\multicolumn{3}{|c|}{\textbf{Application}} & \textbf{NN Types} & \textbf{Compute Type}\\
\textbf{Learning Technique} & \textbf{Domain} & \textbf{Task} & \textbf{Models} & \\
\midrule
\textbf{Supervised} & \textbf{Vision} & Image Classification & MLPs, {\textbf{ResNet}}, VGG, AlexNet, InceptionV3 & FC, CNV \\
& & Object Detection & Faster R-CNN, Yolo9000, Yolov2 & FC, CNV \\
& & Semantic Segmentation & Mask R-CNN, SSD & FC, CNV \\\midrule
& \textbf{NLP} & Machine Translation & Transformer, Seq2Seq & FC, CNV, recurrent \\
& & Speech Recognition & {DeepSpeech2} & FC, CNV, recurrent \\
& & Sentiment Analysis& Seq-CNN & FC, CNV, recurrent \\
& & Language Modeling & Memory Networks & memory network \\
& \textbf{Recommendation} & Movies & NCF & ...\\\midrule
\textbf{Unsupervised} & \textbf{Vision} & Feature Extraction & {Autoencoder} & FC \\\midrule
\textbf{Generative Adversarial Learning} & \textbf{Vision} & Image Generation/Modification & WGAN & NV,DCNV \\\midrule
\textbf{Deep Reinforcement Learning} & \textbf{Game} & Go & MiniGo & ...\\
& & Atari ALE & DeepQ, A3C & ... \\
\bottomrule
\end{tabular}
}
  \vspace*{-3ex}
\end{table}

Table \ref{table:nnoverview} shows the pool of candidate neural networks that we plan to use as part of our benchmark, including both inference and training.
While there is a large breadth of neural networks, there are many common layer types being used, which are ideal to form levels 1 and 2 of QuTiBench.
These layer types equate to the basic computational patterns and are based on previous analysis~\cite{fathom}. The most popular compute layers are \emph{fully connected, convolutional, pooling, normalization} and \emph{recurrent layers}. These come with very different compute and memory requirements and are briefly discussed here.
A more detailed description can be found in \cite{sze2017efficient}.
Fully connected layers compute the full cross product between input tensors (for example) and a vector of weights, the latter are determined during training. Summed to a bias, this is then fed into  an \emph{activation function}.  
Popular activation functions include
the hyperbolic tangent function and the 
rectified linear unit (ReLU). 
In convolutional layers, the output receives inputs from  a small \emph{receptive field} of the previous layer. This approach greatly reduces the number of parameters (or weights) involved and allows local features (e.g., edges, corners) to be found~\cite{mnist}.
A basic 2D convolutional layer is similar to a fully connected layer except that:
a) each neuron receives an image as input and produces an image as its output (instead of a scalar);
b) each synapse learns a small array of weights which is the size of the convolutional window; and
c) each pixel in the output image is created by the sum of the convolutions between all synapse weights and the corresponding images.
Recurrent layers are characterized by the fact that they contain state over a sequence of input data. 
There are many different options for the implementation of the recurrence within the layer, starting from simple recurrent layers, to GRUs or LSTM layers, which can be uni- or bidirectional, feature different numbers of feedback gates, and may include numerous specializations such as peepholes and CRCs.
Beyond, these basic layer types, there are many layer combinations emerging, such as inception layers in GoogleNet\cite{googlenet,inception}, residual layers in ResNet models \cite{resnet}, and so-called fire modules \cite{squeezenet}.
During training using backpropagation with stochistic gradient descent, we need to compute the relative derivative to all inputs for these layers. 
%As is nicely explained in \cite{sze2017efficient} 
This works out to be similar in compute patterns to inference with transposed versions of the inputs whereby significantly larger amount of compute and memory is required~\cite{sze2017efficient}.
However, additional compute such as batch normalization needs to be addressed.

\subsection{Optimization Techniques}
As mentioned in the introduction, 
the challenge lies within the compute and memory requirements which can often preclude inference deployment within the IoT context. 
To alleviate the computational burden and maximize performance, many optimization techniques have been introduced. Particularly successful techniques include pruning, compression, low rank approximations and quantization \cite{han2015deep}. 
We discuss quantization, a specific focus of this work, and pruning in more detail below.
All of these techniques fall under the category of algorithmic optimizations. A representative benchmark supports and measures these, as they are essential for viable deployment solutions.

{\bf Quantization \& Numerical Representations}
Transprecision computing is making strides in many application domains \cite{tagliavini2018transprecision,malossi2018transprecision}, and is highly effective for neural network inference. In particular, quantization to reduced precision datatypes, including 8 bit fixed point integer and below, as well as custom floating point formats.  For example, quantized neural networks (QNNs)
have been shown to work extremely well. 
On smaller image classification benchmarks such
as MNIST, SVHN and CIFAR-10,  QNNs achieve state of the art accuracy despite reduction in precision \cite{courbariaux:2016,zhou2016dorefa}, even for partial or full binarization of fully connected and convolutional layers.
XNOR-Net~\cite{rastegari2016xnor} applies convolutional BNNs on the ImageNet dataset with topologies inspired by AlexNet, ResNet and GoogLeNet, report top-1 accuracies of up to 51.2\% for full binarization and 65.5\% for partial binarization, while for the more challenging ImageNet benchmark, there is a small but noticable accuracy drop. 
The resulting solution can run significantly faster in hardware and might still pose an attractive design trade-off.
Furthermore, there is significant evidence that increasing network layer size can recuperate this drop in accuracy~\cite{fraser2017scaling, sung:2015, wideresnet, mishra2017wrpn, kim:2016}.

\begin{wraptable}{r}{9cm}
  \caption{Latest Accuracy of QNNs}
%\begin{tabular}
  \label{tabHWGQaccuracy}
%  \resizebox{0.5\width}{!}{
  \begin{tabular}{|l|cc|}\toprule
    Network & float top-1(top-5) & QNN top-1(top-5)\\\midrule
    %   ImageNet & AlexNet & 58.5\% (81.5\%) & 52.7\% (76.3\%) \\
    GoogLeNet & 71.4\% (90.5\%) & 63.0\% (84.9\%) \\
    VGG-like & 69.8\% (89.3\%) & 64.1\% (85.6\%) \\
    ResNet-50~\cite{alemdar2017ternary, zhu2016trained} & 79.26 (94.75\%) & 64.6\% (85.9\%) \\
    ResNet-50~\cite{zhuang2017s} && 64.6\% (87.8\%) \\
    \bottomrule
%  \end{tabular}}
%  \vspace*{-3ex}
\end{tabular}
\end{wraptable}

New quantization schemes show promising results using for example
Half-wave Gaussian Quantization (HWGQ)~\cite{cai:2017} to take advantage of the
Gaussian-like distribution of batch normalized activations.
Furthermore, new training and optimization techniques \cite{mishra2017apprentice, zhuang2017s} work effectively.
The current lowest error rates for ImageNet classification have been achieved using ternarization~\cite{alemdar2017ternary, zhu2016trained} as shown in Table \ref{tabHWGQaccuracy}. 
Quantization has  been successfully applied to other tasks including 3D object recognition, facial expression recognition \cite{ma2017bv, sun2017efficient}, optical character recognition as well as speech~\cite{lu2017toward, han2017ese,finnl}.
Even in training, research shows that 32bits are not really needed given the typical value ranges for weight and activation gradients and weight updates involved. Fixed point integers, half precision floating point (FP16), bfloat16, flexpoint or block floating point representations show state-of-the-art performance~\cite{wage,flexpoint,nvidiahp,gupta2015deep}.
All of these need to be accurately reflected within the tests.

{\bf Pruning} This is another popular optimization which has been shown to dramatically reduce memory requirements, through either synaptic pruning or filter pruning. When synaptic pruning is leveraged, irregular compute patterns result which impact  memory access efficiency, thus hardware architectures require support for sparse matrix representations to benefit from this~\cite{han2017ese}. Filter pruning yields regular compute patterns and benefits thereby a broader selection of platforms~\cite{han2015deep}.

 %Ce
\section{Neural Networks and their Compute and Memory Requirements}
\label{sec:cmpmem}

We analyze neural networks with regards to their arithmetic compute, intermediate storage requirement and memory footprint. 
While actual hardware requirements depend on numerous attributes, at this point we are  characterizing the theoretical requirements in an architecturally independent way. For example, actual on-chip requirements and external memory requirements depend on implementation choices, but can be derived directly, so this analysis is useful to categorize the different requirements.
The scope of the analysis is currently constrained to the models shown in Figures~\ref{fig:crt}; The planned scope is listed in the appendix.

\begin{figure}
\centering
\includegraphics[width=0.7\linewidth]{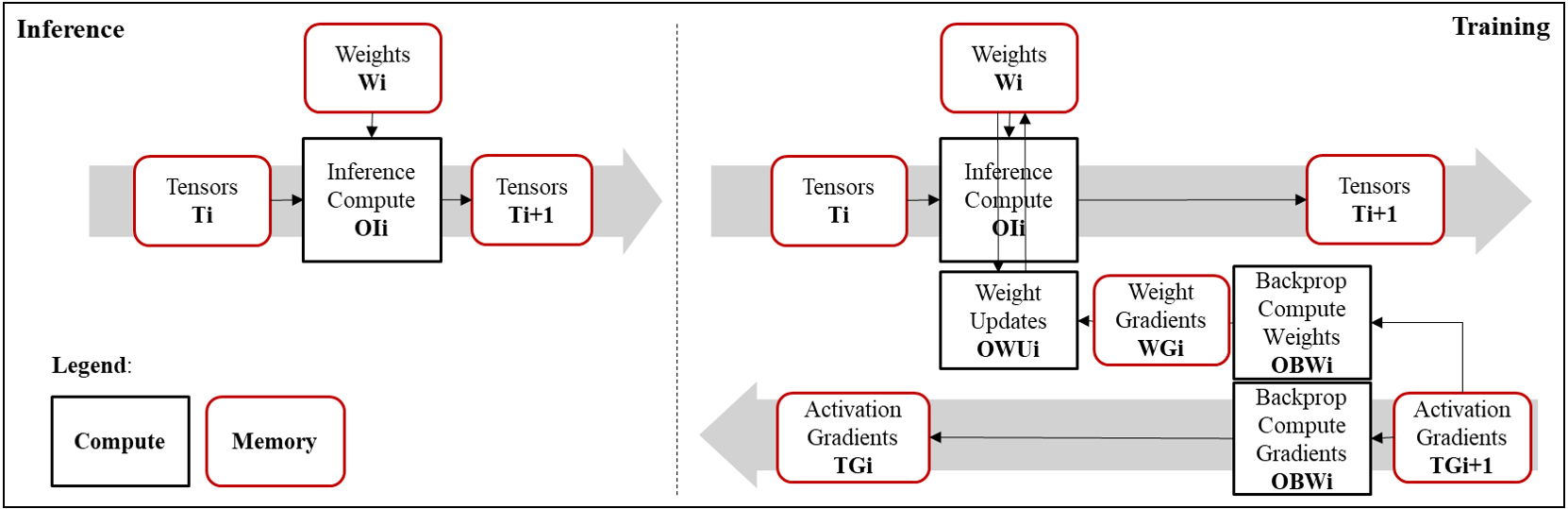}
\caption{Compute, Buffer and Storage Elements}
\label{fig:WLT}
  \vspace*{-3ex}
\end{figure}

{\bf Inference}
Each NN layer (L0, L1, etc.) requires a specific number of arithmetic operations $\mathbf{O_{L0}, O_{L1}, O_{L2}}$ in the form of multiplies, additions etc. We measure these in giga or tera operations respectively (GOPs, TOPs). The overall compute of a network with $\mathbf{n}$ layers,  $\mathbf{O_{total}}$, is the sum of the compute in each individual layer (see eq.~\ref{eq:tens}). 
We define the total modelsize $\mathbf{W_{total}}$ as the sum of the weight requirements per layer measured in millions of elements (ME); this is independent of any choice in numerical representation. The real memory footprint can be derived by multiplying with the size of the given datatype (for example 32b for single precision floating point).
We quantify the intermediate buffer requirement $\mathbf{T_{total}}$ in an implementation neutral fashion. For this we calculate the sum of the required amount of tensors $\mathbf{T_i}$ that precede each layer. These are derived as the product of feature map dimensions ($\mathbf{w_i, h_i}$) and number of channels ($\mathbf{ch_i}$).  Note that all of this applies to non-linear topologies such as DenseNet \cite{densenet}; however, our models currently do not reflect graph connectivity. We plan to address this in the future.

\begin{equation}
O_{total} = \sum_{i=0}^{n-1}O_{i}, \quad W_{total} = \sum_{i=0}^{n-1}W_{i},  
\quad T_{total} = \sum_{i=0}^{n-1} T_{i}, T_i = w_i \times h_i \times ch_i  \\ 
\label{eq:tens}
\end{equation}

{\bf Training}
While training is currently the focus in the cloud, we expect that it will become essential in embedded as well as on-line learning takes off.
In regards to requirements, we need to consider backpropagation in addition to inference. As depicted in Figure~\ref{fig:WLT}, training requires additional data structures. First of all, symmetrically to the tensors  $\mathbf{T_i}$, we need to buffer their gradients $\mathbf{TG_i}$. Furthermore, so-called weight gradients need to be stored $\mathbf{WG_i}$ which are the derivative (in relation to the input weights) of the gradient $\mathbf{TG_i+1}$. Depending on given optimization strategies, weight updates need to be buffered as well. This results in roughly 3 times the buffer requirements for weights, and double the amount for tensors. Regarding compute, backpropagation requires roughly 3 times the inference compute for a single image of the training data set (plus 1 update operation per weight parameter). Overall compute needs to be multiplied with number of iterations and number of inputs in the training data set.
Note that data dependencies are significantly more intricate and challenging for training. This is currently not reflected within the theoretical analysis.

\begin{figure}
\centering
\includegraphics[width=\linewidth]{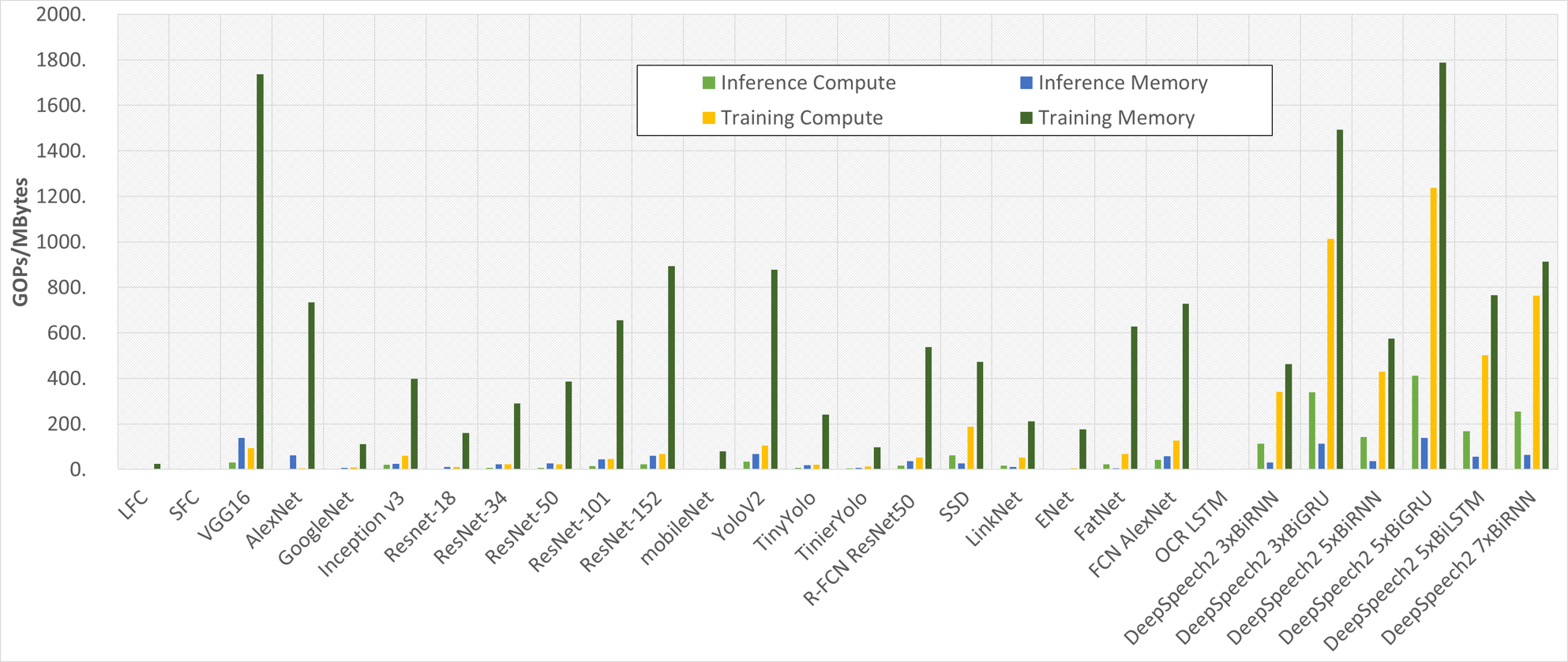}
\caption{Compute Requirements Training and Inference for a spectrum of NNs (Visualization of QuTiBench Level 0 - CNN Statistics)}
\label{fig:crt}
\end{figure}

{\bf Summary of Requirements}
Figure~\ref{fig:crt} visualize initial results, where for Seq2Seq models, we assume a sequence length of 3000 (based on the LSTM test case in DeepBench \cite{deepbench}).
The key observations are as follows: First, the compute and memory requirements are on average very high. Mean model size is too big to fit into most on-chip low latency memory (with 71.14MBytes), and compute is in the GOPs range for every single input datum. Second, there is a significant variation in all requirements for both training and inference as summarized in Table~\ref{table:mean}.
No simple generalizations can be made, even within subcategories such as image recognition, as models vary greatly depending on size and complexity of images, number of objects to be recognized, etc.   
The defined parameters: $\mathbf{O_{total}}$, $\mathbf{W_{total}}$, $\mathbf{T_{total}}$, $\mathbf{OT_{total}}$, $\mathbf{WU_{total}}$, ~and $\mathbf{TG_{total}}$ help describe the compute requirement for inference and training of each individual network and can be used for baseline computations, taking architectural constraints into consideration, and cross-correlated with roofline models to provide rough performance guidance.

{\small
\begin{table}
  \caption{Ranges and Mean Requirements}
  \label{table:mean}
  \resizebox{0.7\width}{!}{
  \begin{tabular}{|l|ccc|ccc|}\toprule
 & \multicolumn{3}{c|}{\textbf{Inference}} & \multicolumn{3}{c|}{\textbf{Training}} \\
 & $\mathbf{OI_{total}}$ [GOPs] & $\mathbf{W_{total}}$ [MBytes] & $\mathbf{T_{total}}$ [MBytes] 
 &  $\mathbf{OT_{total}}$ [GOPs] & $\mathbf{WU_{total}}$ [MBytes] & $\mathbf{TG_{total}}$ [MBytes]\\\midrule
\textbf{Min} & 0.00 & 0.00 & 0.13 & 0.00 & 0.27 & 0.00\\
\textbf{Max} & 412.17 & 71.14 & 138.34 & 1236.64 & 276.69& 71.14\\ 
\textbf{Mean} & 62.59 & 11.9 & 38.02 &  187.79 & 76.05 & 11.9 \\
\bottomrule
\multicolumn{7}{l}{Assuming 8b datatypes for inference and 32b for training.} \\

  \end{tabular}
}
\vspace*{-3ex}
\end{table}}

 %Michaela
\section{Hardware Architectures for Deep Learning}
\label{sec:hw}

\begin{comment}
\begin{figure}
\centering
\includegraphics[width=0.7\linewidth]{hwArchCmp.png}
\caption{Hardware Architectures}
\label{fig:bpfull}
\end{figure}
  \vspace*{-3ex}
\end{comment}

{\small
\begin{table}
\caption{Hardware Architectures for Cloud Systems with Theoretical Performance (QuTiBench Level 0 - Hardware Platform Statistics)}
\label{tableHW-cloud}
\begin{center}
\resizebox{\textwidth}{!}{
\begin{tabular}{|l|ccccc|}
\toprule
\textbf{Platform}&\textbf{Num. Choice}& \textbf{Throughput [TOPs]}&\textbf{Mem BW [GBps]} & \textbf{Power [Watt]}&\textbf{Performance/Power [TOPs/Watt]}\\\midrule
\rowcolor[gray]{.9} \textbf{GPUs} & & & & & \\
NVIDIA V100~\cite{NVIDIAv100} & FP32 & 14 & 250 & 300 & 0.06 \\
NVIDIA V100 & FP16 & 112 & 250 & 300 & 0.45 \\
NVIDIA P100~\cite{NVIDIAp100} & FP32 &   8 & 732 & & \\
NVIDIA P100 & FP16 & 16 & 732  & & \\
NVIDIA P40~\cite{NVIDIAp40} & INT8 & 47 & 200 & 346 & 0.24 \\
NVIDIA P4  & INT8 & 22 & 60 & 192 & 0.37 \\
AMD Vega10~\cite{AMDVega} & FP32 & 13.7 & 484 & 345 & 0.04\\\midrule
\rowcolor[gray]{.9} \textbf{TPUs} & & & & & \\
Google TPUv1~\cite{tpu1} & INT8 & 92 & 75 & 34 & 1.23 \\
Google TPUv2~\cite{tpu} & FP16 & 45 &  & 600 &  \\
Google TPUv3~\cite{tpu3} & FP16 & 90 &  &  &  \\\midrule
\rowcolor[gray]{.9} \textbf{ASIC DPU} & & & & & \\
Graphcore & Custom & 224 & & 300 & 0.75 \\
Groq & unknown & 400 & & & 8 \\
Nervana & custom16 & 55 & & & \\
Wavecomputing 1DPU & INT8 & 181 & & 271 & 0.7 \\\midrule
\rowcolor[gray]{.9} \textbf{FPGA DPU} & & & & & \\
Xilinx VU9P	&2b/8b	&93.00	&88	&100	&1.06\\
Xilinx VU9P	&2b/4b	&139.88	&88	&100	&1.59\\
Xilinx VU9P	&2b/2b	&192.52	&88	&100	&2.19\\
Microsoft Brainwave Stratix X~\cite{chung2018serving} & FP8 & 90 & & 125 & 0.72  \\
\bottomrule
\end{tabular}
} 
\end{center}
  \vspace*{-3ex}
\end{table}}

{\small
\begin{table}
\caption{Low Power Hardware Architectures and Theoretical Performance (QuTiBench Level 0 - Hardware Platform Statistics)}
\label{tableHW-embedded}
\begin{center}
\resizebox{\textwidth}{!}{
\begin{tabular}{|l|ccccc|}
\toprule
\textbf{Platform}&\textbf{Num. Choice}& \textbf{Throughput [TOPs]}&\textbf{Mem BW [GBps]} & \textbf{Power [Watt]}&\textbf{Performance/Power [TOPs/Watt]}\\
\midrule
\rowcolor[gray]{.9} \textbf{CPUs} & & & & & \\
Bitserial Cortex-A57 on Jetson TX1~\cite{umuroglu2017streamlined} & BIN & 0.09 &  &  & 0.019 \\\midrule
\rowcolor[gray]{.9} \textbf{GPUs} & & & & & \\
NVIDIA TX2 (MaxP)~\cite{nvidia-jetson} & FP32 & .575 & 59.7 & 15.0 & 0.038\\
NVIDIA TX2 (MaxP)~\cite{nvidia-jetson} & FP16 & 1.15 & 59.7 & 15.0 & 0.077\\\midrule
\rowcolor[gray]{.9} \textbf{ASIC DPU} & & & & & \\
Movidius Myriad 2~\cite{movidius-tom} &  INT8 & .15  &  & 1.2 & 0.125 \\
Movidius Myriad X~\cite{myriadx1} &  INT8 &  1 &  & 1 & 1 \\
Kalray MPPA Turbocard3~\cite{kalray} & FP32 & 1.6 &  & 110 & 0.014 \\
BinarEye~\cite{binareye} & BIN & 0.09 - 2.8$^\star$ & & & 230$^\dagger$\\
BNN Custom Fabric~\cite{ando2017brein} & BIN & 1.4 & & 0.6 & 2.3\\
Stripes Bitserial ASIC~\cite{judd2016stripes} & BIN & 128.5 & & & 4.3\\
IBM AI Accelerator~\cite{IBMAI}\footnote{We will add the VLSI reference as soon as it becomes live} & BIN & 12 & & & \\
Eyeriss~\cite{chen2017eyeriss} & INT16 & 0.084 & & 1.17 & $^\dagger$\\
ARM ML Processor~\cite{trillium} & unknown & 4.6 &  &  & 3 \\
DianNao~\cite{diannao} & INT16 & 0.452 & 120 & 0.485 & 0.93\\
EIE(28nm)~\cite{han2016eie} & INT4 & 3 (0.102 sparse) & 2.36 & 1.27 & 2.4 (0.08 sparse)\\
Cambricon-X~\cite{zhang2016cambricon}  & INT16 & 0.544 &  &  & \\
\midrule
\rowcolor[gray]{.9} \textbf{FPGA DPU} & & & & & \\
Lattice SenseAI~\cite{lattice-bnn}  & BIN & 1.4 & & 0.6 & 2.3\\
Bismo biserial on PYNQ~\cite{umuroglu2018bismo}& BIN & 6.5 &  & 4.64 & 1.4 \\
FINN on ZC706~\cite{umuroglu2017finn} & BIN & 11.6 & & & 0.408 \\
ZCU104 (Deephi-666MHz)      & INT8 & 4.60 & 19.2 & %12 
& \\ %0.38 \\
ZCU104 (Theoretical-775MHz) & INT8 & 5.36 & 19.2 & %12 
& \\ %0.45 \\
GX1150 on HARPv2~\cite{moss2018customizable} & BIN & 0.041 & & & 0.85 \\
\midrule
\multicolumn{6}{l}{Measured $^\star$} \\
\multicolumn{6}{l}{Chip level power consumption only $^\dagger$} \\
\end{tabular}
} 
\end{center}
  \vspace*{-3ex}
\end{table}}

We discuss target hardware systems, their architectures and implementation alternatives. While we present details on  cloud platforms, the focus of this article is on embedded systems.  
There is a huge range in the types of hardware architectures used for machine learning applications, including CPUs, GPUs, FPGAs and specialized architectures. The field has spawned significant new research in computer architecture and created so-called deep learning processing units (DPUs), which are specialized for this application domain and can be implemented either with ASICs or in FPGAs.  Architectures can broadly be classified by the basic type of compute operation, memory bandwidth, level of parallelism, degree of specialization and inherent precision support.  CPUs are widely used for ML applications, and are viewed as serial compute engines, optimized for single thread performance, with implicitly managed memory hierarchies (including three levels of caches), and support floating point operations.  GPUs are vector processors that support smaller floating point formats (FP16) natively, most recently fixed point 8bit integer formats, and have a mix of implicitly and explicitly managed memory. DPUs, such as Google's Tensor Processing Unit (TPU), work with tensors, have explicitly managed and specialized memory hierarchies and support integer operations.  With newer generations, the boundaries between different hardware architectures are blurring. CPUs are usually multicore to support parallel processing, and incorporate vector processing units, GPUs are adding tensor processing units, and the TPU now supports floating point operations.
FPGAs can support any of the above configurations with explicitly managed memory.  FPGAs are the most flexible of all target hardware, and can be configured to support any numeric representation,
even bit-serial hardware architectures which provide run-time configurable precision. 
Custom ASIC implementations, which minimize hardware cost and maximize performance, have emerged to exploit specific precision arithmetic and customized memory systems.
Tables~\ref{tableHW-cloud} and \ref{tableHW-embedded} list many of these hardware targets along with published performance numbers. 
\footnote{These tables form part of level 0 of our benchmark suite and can be used as a basis for performance estimation.}
One of the goals of QuTiBench is to provide a more systematic way to compare performance and accuracy between these systems, rather than relying on vendor reported metrics.  

NVIDIA GPUs are some of the most popular hardware targets for machine learning, and newer families of chips have been introduced to specifically accelerate this task.  For example, the Volta architecture, introduced in 2018, was particularly designed to  accelerate AI and incorporates tensor cores as a new feature, as well as improved FP32 and FP64 support for training in a data center setting~\cite{NVIDIAv100}. AMD announced the Vega GPU~\cite{RadeonInstinctGPU} with new deep learning instruction set operations, with the goal of obtaining parity with NVIDIA's high-end Tesla V100 datacenter GPUs.  Both companies have low power GPUs: the AMD Vega mobile GPU~\cite{radeon-mobile} and NVIDIA Jetson TX2~\cite{nvidia-jetson}.

Google introduced its TPU in 2016~\cite{tpu1}, which was designed to accelerate Google's TensorFlow framework.  The first generation supported integer arithmetic with a massively parallel 8-bit matix multiply engine.  The second generation TPU was anounced in May 2017~\cite{tpu}, and the third generation in May 2018~\cite{tpu3}.  These newer chips boast improved memory performance as well as support for floating point specifically aimed at training.  

There are a number of startups introducing custom hardware in this space. Within the cloud space, there are Graphcore, Cerebras, Groq, and Wave Computing. Within the embedded space, where the design constraints are even more stringent, we find even more, as are listed in table~\ref{tableHW-embedded}.
Most are secretive about the details of their designs, and this landscape is rapidly changing.  Intel is investigating several custom accelerators including Nervana and Movidius. 
Fathom~\cite{movidius-tom} is Movidius' ultra low power Neural Compute Stick which operates at about 1 Watt.  
At the extreme, binarized neural networks which are very high throughput at extremely low power,  are exploited in the following ASICs: BinarEye~\cite{binareye}, BNN Custom Fabric~\cite{ando2017brein}, Stripes Bitserial ASIC~\cite{judd2016stripes}, and IBM AI Accelerator~\cite{IBMAI}. Others exploit sparse computing engines, such as EIE and its successor ESE~\cite{han2017ese}, SCNN~\cite{parashar2017scnn}, Cnvlutin~\cite{cnvlutin}, Cambricon-S and Cambricon-X~\cite{zhang2016cambricon}.

FPGAs are an extremely popular platform for machine learning.  As they are highly flexible and can be used in a variety of different configurations and support any arithmetic format, they can be fully customized towards specific neural network topologies, thereby achieving high performance and efficiency.  However, for the same reason, they are  extremely difficult to characterize in general.  FPGAs are available in the cloud,  such as the Xilinx Ultrascale+ VU9P available as part of the public Amazon Web Services (AWS) cloud infrastructure. 
Within the embedded space, we have pioneered the first binarized neural network accelerators~\cite{umuroglu2017finn,fraser2017scaling} and provided many proof points for customized reduced precision implementations~\cite{iccd}.  Umuroglu et al.~\cite{umuroglu2018bismo} demonstrates that run-time programmable precision can be achieved with a bitserial approach, providing highly attractive performance on FPGAs, with little overhead.
Intel FPGAs have also been successfully applied to machine learning applications using a range of different numerical representations~\cite{Nurvitadhi2017CanFB}.  
The Microsoft Brainwave project~\cite{chung2018serving} aims at applying FPGAs at datacenter scale using their own custom floating point representation.  
Focusing on the IoT market, Lattice has announced binarized neural network libraries targetting low power FPGAs and achieving 1TOPS/Watt~\cite{lattice-bnn}.  
 %Miriam
\section{Characteristics \& Challenges in Benchmarking}
\label{sec:whybench}

\subsection{Key Components of a Benchmark}

\begin{wrapfigure}{r}{0.6\linewidth}
  \begin{center}
   \includegraphics[width=0.6\textwidth]{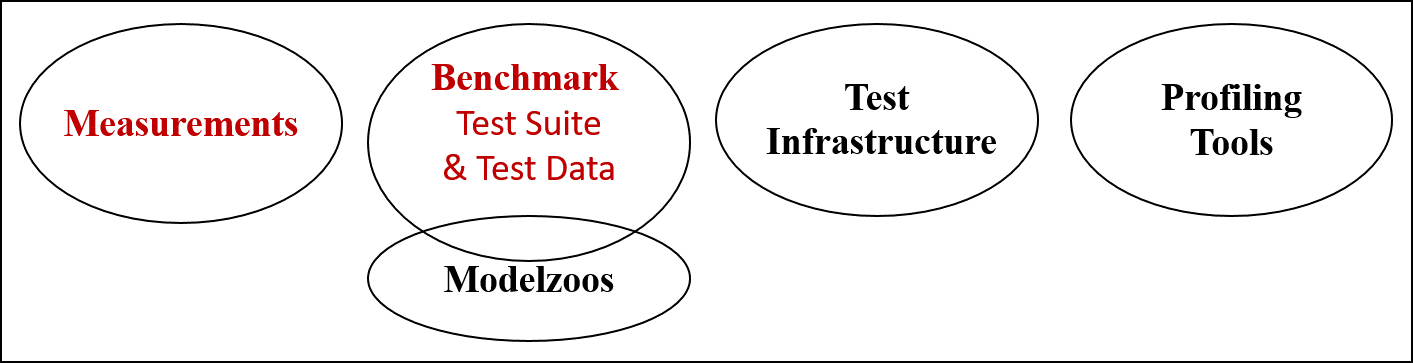}
  \end{center}
  \caption{Benchmarking collateral}
  \label{fig:col}
\vspace*{-2ex}
\end{wrapfigure}

A benchmark can be defined as a set of standards used for evaluating performance or level of quality. A more practical definition implies that the ``set of standards'' is supplied in the form of a well-defined set of executable tests and measured regarding a specific set of figures of merit. 
%The tests combined with the defined figures of merit are the essence of the benchmark. 
Sometimes additional items are included such as performance analysis or profiling tools which can help shed light on system bottlenecks. Test infrastructure or a testbed can be provided to ensure reproducibility. This makes particular sense when specialized and not easily available hardware systems are involved.  Data management can be handled together with the benchmark suite and stored in an accessible location as for example with DAWNbench~\cite{dawnbench}, MIT's Eyeriss project~\cite{eyeriss-benchmark} and the Request tournaments online score card~\cite{request}. 
In this article we differentiate profiling tools, test infrastructure, and measurements from the actual benchmark test suite (see Fig.~\ref{fig:col}).
Somewhat related to benchmarking are modelzoos, such as OpenAI Gym \cite{brockman2016openai} and rllab \cite{rllab}, which are selections of sample code. They are not necessarily aiming to be representative, and typically include simplified implementations to teach concepts.
QuTiBench focuses initially on the benchmark suite and measurements.

\subsection{Characteristics}

Benchmarking can bring many insights. For end-users and system designers, it helps to estimate expected system-level performance and provides an understanding of what algorithms work best on which hardware platform. 
For hardware designers, benchmarks provide design perspectives and clear cut guidelines regarding what figures of merit matter and what workloads look like.  Neural networks are pushing the limits of what is possible, therefore careful system level co-design of hardware and algorithms, and realistic expectations of what is achievable given the design choices using benchmarking, are crucial.
To bring maximum benefit, the following characteristics are essential which are discussed in greater detail below:

\begin{itemize}
\item representative of common workloads
\item supportive of algorithmic modifications
\item objective and reproducible
\item portable to heterogeneous hardware systems
\item complexity vs accuracy tradeoff
\item adaptive ``living" benchmark supported by industry and academia
\end{itemize}

{\bf Representative}
Benchmarks need to be \emph{representative} of real world workloads. In machine learning, 
this requires breadth across a spectrum of applications, algorithms and computational patterns. 
Computational patterns are important to maximize insights into different hardware architectures. Application coverage is essential as it provides more holistic insights into system level performance which can be hard to predict given the emerging complexity of increasingly heterogeneous hardware systems. 

{\bf Support for algorithmic modification}  Algorithmic modifications are inevitable to extract best possible performance out of diverse hardware systems, for example to take advantage of caching and parallel hardware resources. 
Within machine learning, software and hardware co-design are compulsory \cite{guo2017software} for energy constrained compute environments.  To support this algorithmic freedom within the benchmark suite, 
application coverage is essential,  as we correlate hardware performance independent of the algorithm back to application performance, which is equivalent to accuracy in this context.
However, optimized performance alone is not sufficient, as not every system designer may be able to achieve it. 
We also need to reflect the out-of-the-box, naive performance. Both optimized and naive are representative of a specific hardware platform, and the difference gives a good indication of the development effort involved.  We believe both should be part of the benchmarks and be captured together with development time or lines of code.
Specifically for neural networks, quantization, compression, topological changes and pruning techniques are important optimization techniques that need to be considered.

{\bf Objective \& Reproducible}
To provide clear differentiation between marketing and scientific efforts, reproducible and objective results that do not favour any particular system configuration or hardware architecture are needed.
\emph{Reproducible} results are a key ingredient in the move towards \emph{Open Science}, however, what does reproducibility actually entail?
In the context of the plethora of esoteric AI accelerators, is it sufficient that an objective third party has validated the results? %as is for example done in the context of the Collective Knowledge Framework \cite{ckframework}? 
Or does it imply that everyone on the planet should be in a position to reproduce the results if they had access to the system at a reasonable cost? 
Some  hardware systems are too expensive; for example, a NVIDIA V100 may be beyond someone's budget. Other hardware choices are only available for rent, such as Google's TPU versions as part of Google cloud. 

{\bf Portability}
is a challenging subject as specialized hardware architectures come with their own design entry languages and compiler tool stacks.
The community is fragmented by a huge choice of frameworks including Caffe, Tensorflow, Mxnet, Theano, pytorch and Darknet. What is more, the prediction accuracy of a network depends on the choice of framework,  since training data is passed through different preprocessing stages and numerical inaccuracies accumulate and manifest themselves as discrepancies. These inaccuracies are exacerbated by the characteristics of floating point arithmetic~\cite{gu2015behavioral}.  As a result, models and frameworks are inherently tied together.
There are three basic choices: The first is to constrain ourselves to exactly one framework as was done with  Fathom~\cite{fathom}. Second, we could support all frameworks. However, given that we are dealing with different hardware backends, this causes an explosion in test infrastructure, as the number of tests multiplies with the number of frameworks.
The final choice and probably the cleanest, is to support one of the intermediate neural network representations such as ONNX~\cite{onnx}, NNEF~\cite{nnef} or TVM~\cite{tvm}, which provide translation between all popular frameworks. However, this requires hardware vendor support, which is currently limited.  

{\bf Complexity vs Speed vs Accuracy}
Speed of result is essential, as the key purpose of a benchmark is to provide faster insights than developing the full end-system. There is a trade-off between speed,  benchmark complexity and the accuracy of the results. 
Benchmarks which provide application and algorithmic breadth may require a large number of tests thus making the benchmark suite inherently complex and limit the usefulness of the benchmark. 
Sometimes it is important to have less accurate predictions at a faster rate, and, for different users, different tradeoffs are acceptable.

{\bf Adaptive}
As machine learning is a highly active research field where algorithms change fast, the benchmark suite should be adaptive and able to incorporate emerging popular algorithms, compute patterns and end applications. 
 %Michaela
\section{Related Work: Existing Benchmarking}
\label{sec:rel}

In this section we take a look at existing benchmarks, and compare them regarding algorithmic scope and figures of merit. QuTiBench differs from these efforts in a number of ways: 

\noindent $\bullet$ Existing benchmarks do not address the fact that heterogeneous hardware platforms typically require co-designed algorithms, and offer flexibility in precision for datatypes specifically, although MLPerf has open models  for training. We introduce correlation of application and architecture figures of merit to compare different combinations of algorithms and architectures at the application level. 
\noindent $\bullet$ We offer full visualization of the design space, rather than comparing performance for fixed levels of accuracy. Thus, interesting trade-offs can be highlighted.
\noindent $\bullet$ None of the existing benchmarks offer the some level of tiering, including theoretical level, and stacks of microbenchmarks that can help isolate problematic data movement patterns and tensor dimensionalities.
\noindent $\bullet$ Finally, there is a difference in scope. Most benchmarks currently focus foremost on training. 

\begin{figure}
\centering
\includegraphics[width=0.7\linewidth]{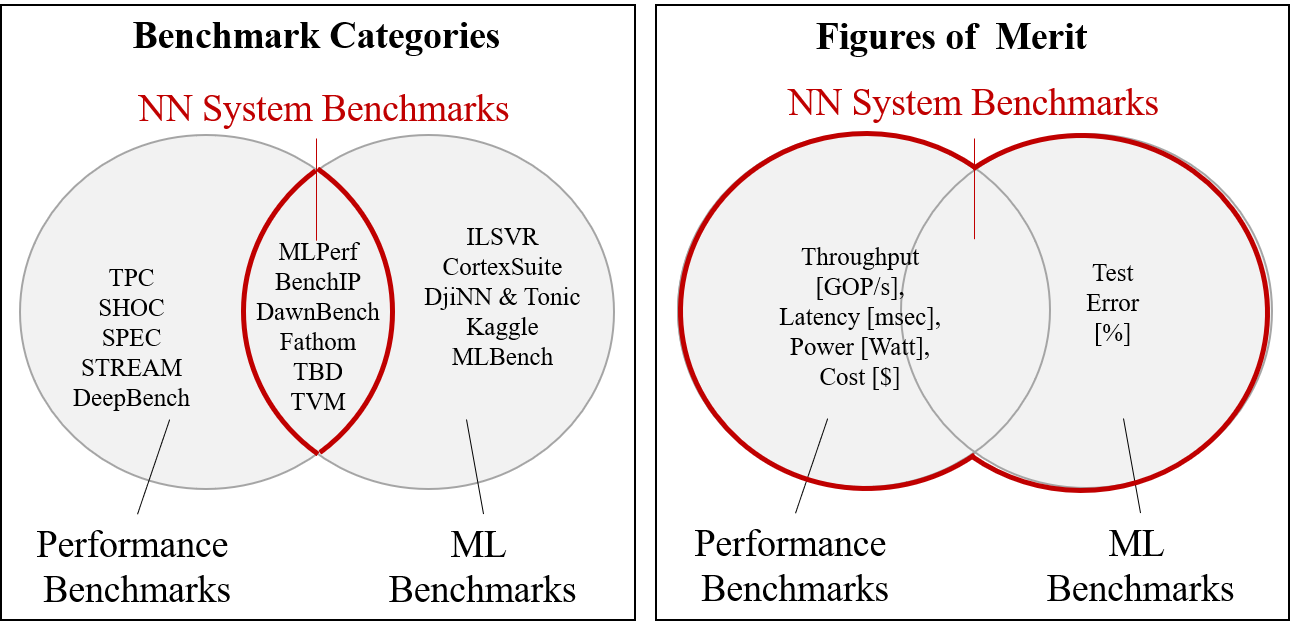}
\caption{Categories of Benchmarks and corresponding Figures or Merit}
\label{fig:nnsystems}
  \vspace*{-3ex}
\end{figure}

In the following, we expand and elaborate on the differences in greater detail.
For this, we differentiate between \textbf{ML benchmarks}, \textbf{performance benchmarks} and \textbf{NN system benchmarks}. ML benchmarks exclusively focus on application performance, which is accuracy. There is no consideration of compute effort required or resulting execution time. Performance benchmarks record hardware performance only, specifically throughput (measured in processed inputs per time or TOP/s), latency or response time in milliseconds (ms), and power consumption in Watts.
Performance benchmarks only look at hardware performance and are agnostic of the application. NN system benchmarks, as shown in Figure \ref{fig:nnsystems} lie at the intersection and are at the heart of what we are striving for. They combine all figures of merit; both system performance and accuracy are correlated. In addition, functional correctness even during performance testing needs to be ensured. 

\subsection{NN System Benchmarks}\label{secR}
QuTiBench falls into this family of benchmarking suites which are unique in that they combine representative machine learning workloads with figures of merit from  hardware performance benchmarks.
\textbf{BenchIP}~\cite{tao2018b} is a benchmarking suite which has a broad set of machine learning tasks. Similar to QuTiBench, BenchIP adopts a multi-tiered approach with micro- and macro-benchmarks. However BenchIP does not support the theoretical layer, which we use to cover compute efficiency and track benchmarking results. BenchIP also doesn't cover level 2, namely stacks of layers, which we believe bring great merit in isolating bottlenecks in data movement and highlighting problematic dimensionality in tensors. Finally BenchIP does not offer the concept of comparison via pareto curves which is essential to a) visualize the full scope of potential solutions within the design spectrum, and b) provide the necessary scope for algorithm optimizations matching the specifics of various accelerators.
\textbf{Fathom}~\cite{fathom} is probably the first attempt to provide a representative workload for benchmarking that has algorithmic breadth beyond convolution neural networks inference and includes example training and unsupervised learning such as reinforcement learning and recurrent models.
However, Fathom does not address the spectrum of numerical representations.  It also does not support heterogeneous hardware platforms. In regards to framework strategy, Fathom advocates a unified software package, relying on compatible software stacks to emerge, 
and therefore only supports one framework, TensorFlow.
With a primary focus on benchmarking for training and achieving application coverage rather than algorithmic breadth, \textbf{TBD}~\cite{tbd} adopts some of the concepts introduced in Fathom. It supports more frameworks and datasets and covers a range of applications, including image classification, machine translation, object detection, speech recognition, adversarial and deep reinforcement learning.
\textbf{MLPerf}~\cite{mlperf} is a promising approach at providing system level benchmarks.
Similarly to Fathom and TBD, it covers a representative range of applications adding sentiment analysis and recommendation as target applications. It currently considers only training but inference is in process.
MLPerf is created by a consortium of industry partners and universities, which should address objectivity criteria. Its key strengths are explicitly defining figures of merit and its strong industrial support.
It provides the concept of open models, which allow for algorithmic optimizations that facilitate performance improvements for specific architectures. However, it does not explicitly support quantization.

\textbf{DAWNBench}~\cite{dawnbench} exclusively looks at ImageNet classification for training and inference. The benchmark sets very clear figures of merit such as ``Time taken to train an image classification model to a top-5 test accuracy of 93\% or greater" and ``Latency required to classify one ImageNet image using a model with a top-5 test accuracy of 93\% or greater" and as such supports the concept of algorithmic optimizations by tying hardware performance to accuracy achieved at the application level but falls short of visualizing the full design space. Finally DAWNBench does not provide further insights beyond the specified figures of merit, and is limited in application scope.

The \textbf{Collective Knowledge Framework}~\cite{ckframework} in conjunction with the ASPLOS Request Tournament~\cite{request}, while narrow in scope (limited to ImageNet Classification inference), opens up the design space for different hardware accelerators,  facilitating architecture specific algorithmic transformations and correlation between accuracy and performance and power within a larger design space. This is essential to support heterogeneous hardware architectures. ASPLOS excels in reproducibility, leveraging ACM artifact evaluation technology, and providing insight into hardware performance and error rate trade-offs, through an online scorecard.

\subsection{ML Benchmarks}
The Machine Learning community has defined its own benchmarks which have an exclusive focus on achieved accuracy independent of the required compute, employing ensemble techniques and multi-crop which in essence, linearly scale up the compute load per input data. The most popular of these is the \textbf{ImageNet Large Scale Visual Recognition (ILSVR) Challenge}~\cite{ILSVRC}. The associated compute requirements are unrealistic, particularly when deployed in energy-constrained environments.
\textbf{CortexSuite}~\cite{thomas2014cortexsuite} and \textbf{BenchNN}~\cite{chen2012benchnn} are limited to measuring accuracy, where CortexSuite is constraint to perception and cognition while BenchNN shows the value of machine learning for approximate computing, based on 5 out of the 12 recognition, mining and synthesis applications from the PARSEC benchmark suite.
\textbf{DjiNN and Tonic}~\cite{djinn} focuses on deep learning tasks for warehouse scale computers including image, speech processing and natural language processing. 
While \textbf{kaggle}(www.kaggle.com) isn't specifically a benchmark, it hosts a portfolio of data science challenges where the machine learning community competes with the latest topologies and algorithms for highest accuracy. \textbf{MLBench}~\cite{mlbench} compares human derived learning algorithms against machine learning services from Amazon and Microsoft Azur.

\subsection{Performance Benchmarks}

\textbf{DeepBench}~\cite{deepbench} is probably the most successful suite of microbenchmarks for neural network performance that measures and compares basic compute operations. It benchmarks individually direct convolutions, 
matrix multiply, and a specific LSTM layer for single precision, half precision floating point and for some operations 8b fixed point integer datatypes on hardware architectures. It currently features cloud deployment and some embedded data points on raspberry pi and iphone. It captures the most popular compute patterns, however lacks support for lower precision datatypes, and exclusively investigates performance. As such it does not provide the mechanisms to tie algorithmic modifications back to the application level, nor provide insights into compute performance for reduced precision representations.
DeepBench also doesn't cover data movement bottlenecks between layers, as well
as potential bottlenecks around buffering state, as required for LSTMs for example,  where capacity and access latency crucially impact overall speed. 

There are more general, machine learning agnostic, hardware benchmarks such as \textbf{TPC}~\cite{tpc} for the data processing community, \textbf{SHOC}~\cite{shoc}, \textbf{SPEC}~\cite{spec} and \textbf{STREAM}~\cite{stream}.
SHOC looks specifically at how to benchmark heterogeneous hardware systems using OpenCL as design entry. Similar to QuTiBench, SHOC deploys microbenchmarks combined with application benchmarks and is multi-tiered. SPEC includes a broad range of applications including graphics, MPI, mail servers, virtualization, and storage, and STREAM exclusively focuses on  memory bandwidth.
None are specifically designed for machine learning, and address the challenges of this application domain.
\textbf{gemmlowp}~\cite{gemmlowp}, while it is not a benchmark, is  specifically designed for matrix multiply operations; it includes low precision operations which may be suitable as a basis for implementation of part of our benchmark suite.

{\bf Summary}
Overall, support for algorithmic optimization is limited across the whole spectrum of benchmarks, in particular in regards to quantization and pruning. None of the benchmarks above provide a multi-tiered approach in the same way we do.  These can provide understanding of compute and data movement bottlenecks within the system, or offer theoretical levels with efficiency tracking. None of the benchmarks offer a fair comparison for co-design algorithms and full design space visualization.
In Tables \ref{tableApps} and \ref{tableBAM}, we summarize the application scope of existing and our proposed benchmark, as well as the key differentiators between existing benchmarks and our proposal and discuss in Sec.~\ref{sec:proposal} how we address these characteristics.

{\small
\begin{table}
\caption{Benchmarks, Applications, Datasets and Models}
\label{tableApps}
\resizebox{\textwidth}{!}{
\begin{tabular}{|l|c|c|c|c|}
\toprule
\textbf{Application} & \textbf{MLPerf}&\textbf{Fathom}&\textbf{TBD}&\textbf{BenchIP} \\
\textbf{Domain - Task} & \textbf{ Dataset - Model} & \textbf{Dataset - Model} & \textbf{Dataset - Model} & \textbf{Dataset - Model}\\\midrule
\rowcolor[gray]{.9} \textbf{Supervised Learning}  & \multicolumn{4}{c|}{} \\\midrule
\textbf{Vision - Image Classification} & ImageNet - ResNet & ImageNet - ResNet & ImageNet1k - ResNet50 & ImageNet - ResNet\\
  & & ImageNet - VGG, AlexNet & ImageNet1k - InceptionV3 & ImageNet - VGG, AlexNet\\
\textbf{Vision - Image Classification} & & & & MNIST - LeNet-5\\
 \textbf{Vision - Object Detection} & COCO  & -  & Pascal VOC 2007 - Faster R-CNN & Pascal VOC 2012 - Faster R-CNN \\
 \textbf{Vision - Semantic Segmentation}  & Mask R-CNN & - & - & Pascal VOC 2012 - DeconvNet \\
\textbf{Vision - Image Captioning} & - & - & - & Visual Gnome - FCLN\\
\textbf{Vision - Video Captioning} & - & - & - & MSVD - S2VT\\
\textbf{Vision -  Face Recognition} & - & - & - & LFW - Deep Face Recog\\ 
   \textbf{NLP - Machine Translation} & WMT Eng-German - Transformer & WMT-15 - Seq2Seq & IWSLT15 - Seq2Seq & English WSJ - SyntxNet\\
  \textbf{NLP - Machine Translation} &  &   & IWSLT15 - Transformer & \\
  \textbf{NLP - Speech Recognition} & {Librispeech} - {DeepSpeech2} & TIMIT - {DeepSpeech} & {Librispeech} - {DeepSpeech2} & RNN - WSJ \\
  \textbf{NLP - Sentiment Analysis} & IMDB - Seq-CNN & - & - & -\\
  \textbf{NLP - Language Modeling} & - & babI - Memory Networks & - & -\\
\textbf{Recommendation - Movies} & MovieLens-20M - NCF & - & - & \\

\midrule
\rowcolor[gray]{.9} \textbf{Unsupervised Learning} & \multicolumn{4}{c|}{} \\\midrule
\textbf{Vision - Feature Extraction} & - & MNIST - Autoencoder & - & - \\
  \textbf{Vision - Adversarial Learning} & - & - & Downsampled ImageNet - WGAN & -\\
  \textbf{Recommendation}  & - & - & - & -\\\midrule
\rowcolor[gray]{.9} \textbf{Deep Reinforcement Learning} & \multicolumn{4}{c|}{} \\\midrule
\textbf{Game - Go} & Go - Mini-Go &   &  & \\
\textbf{Learning - Atari ALE} & Atari ALE - Deep Q & Atari2000 - A3C & \\
\bottomrule
\end{tabular}
}
  \vspace*{-3ex}
\end{table}
}

{\small
\begin{table}
\caption{Feature Comparison of Existing Benchmarks and QuTiBench}
\label{tableBAM}
\resizebox{0.7\width}{!}{
\begin{tabular}{|l|ccccccc|}
\toprule
\textbf{Criteria} & \textbf{MLPerf}& \textbf{DeepBench} & \textbf{DawnBench} & \textbf{Fathom} & \textbf{TBD} & \textbf{BenchIP} & \textbf{QuTiBench}\\ \midrule
\rowcolor[gray]{.9} \textbf{Machine Learning Task} & & & & & & & \\
Training & yes & micro & yes & micro & yes & yes & planned \\
Inference & planned & micro & yes & micro & & yes & yes (Sec.~\ref{sec:experiment})\\
\midrule
\rowcolor[gray]{.9} \textbf{Coverage - see Table~\ref{tableApps}} & & & & & & & \\
Applications    & broad &        & narrow &       & broad &  broad & broad \\
Compute Patters & broad & medium & narrow & broad & broad & broad & broad \\
Data Movements &  &  &  &  &  & & broad \\
\midrule
\rowcolor[gray]{.9} \color{red}\textbf{Support for Algorithmic Optimizations} & limited & & limited & & & & yes (Sec.~\ref{sec:experiment})\\\midrule
\rowcolor[gray]{.9} \color{red}\textbf{Full Design Space Representation} & yes & & yes & & & & yes (Sec.~\ref{sec:experiment})\\
\midrule
\rowcolor[gray]{.9} \textbf{Deployment Target} & & & & & & & \\
Cloud       & yes & yes & yes & yes & yes & yes & planned \\
Embedded    & yes & yes &     & & & yes & yes (Sec.~\ref{sec:experiment})\\
\midrule
\rowcolor[gray]{.9} \color{red}\textbf{Benchmark Abstraction} & & & & & & & \\
Theoretical & & & & & & & yes \\
Microbenchmarks Compute & & yes & & yes & & yes & yes (Sec.~\ref{sec:experiment})\\
Microbenchmarks Data Movement & & & & & & & yes (Sec.~\ref{sec:experiment})\\
Full Applications & yes & & yes & & yes & yes & yes (Sec.~\ref{sec:experiment})\\
\textbf{Speed vs Accuracy Tradeoff} & & & & & & limited & yes (Sec.~\ref{sec:experiment})\\
\textbf{Bottleneck Insights} & & & & & yes & & yes (Sec.~\ref{sec:experiment})\\
\midrule
\rowcolor[gray]{.9} \textbf{Reproducibility} & yes & yes & yes & yes & yes & planned & planned \\
\bottomrule
\end{tabular}
}
  \vspace*{-3ex}
\end{table}
} %Michaela
\section{The Benchmark Proposal}
\label{sec:proposal}
\begin{figure}
\centering
\includegraphics[width=0.8\linewidth]{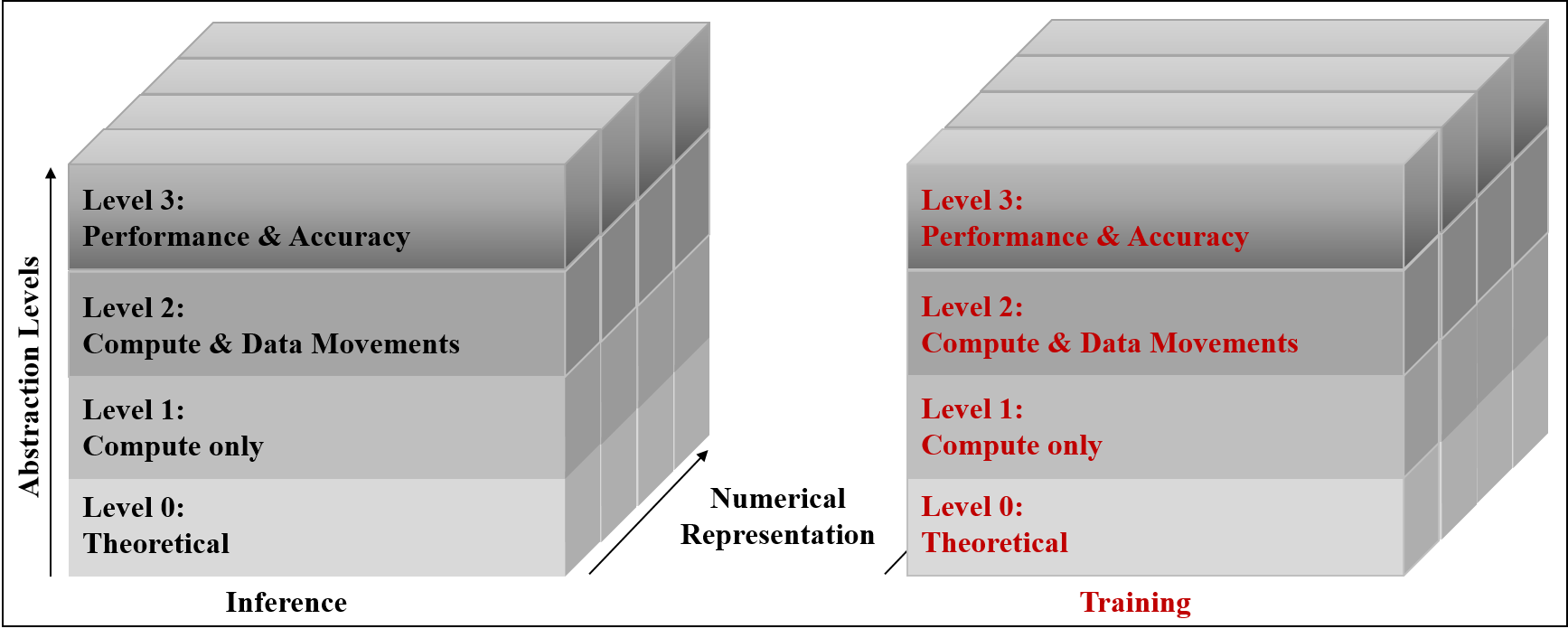}
\caption{A Multi-Layered Approach with Precision Support}
\label{fig:multilayer}
  \vspace*{-2ex}
\end{figure}

The targeted design space is vast and compromised of a multidimensional spectrum of algorithmic and architectural co-designed end solutions.
The aim of the benchmark is to expose the spectrum of possibilities and accurately reflect the capabilities of the different hardware platforms.
QuTiBench has the following key characteristics:
We take a multi-tiered approach which is one of our key contributions (Fig.~\ref{fig:multilayer}). 
We tier the benchmark suite with respect to abstraction levels as well as numerical representations for both training and inference tasks. This provides not only attractive compromises in regards to speed versus minimal discrepancy with target workloads, but also brings advantages such as additional system level insights.

The second key differentiator of our approach is the support for algorithmic optimization by coupling hardware performance with accuracy at the application level. In particular, this allows for objective comparison between floating point implementations and reduced precision models that can achieve much higher performance at a significantly reduced energy cost, among many other possible optimization strategies. Results are visualized via pareto graphs (accuracy versus latency, throughput and throughput/power) and optimal solutions can be found along the pareto frontier.
Third, we include a theoretical level as a baseline for benchmarking and performance estimation.

The unique characteristics of QuTiBench include test suites at various abstraction levels, algorithmic optimizations and quantization, in particular considerations in regards to datasets, hyperparameters and framework challenges, such as reproducibility and adaptibility (see \cite{qutibench}). 

{\bf Multiple Tiers - Abstraction Levels}
\label{sec:tier}
We defined 4 levels of abstraction (Fig.~\ref{fig:multilayer}) discussed below. 

{\bf Level 0 - Theoretical} Records for all target hardware backends theoretically possible peak performance (TOps or GOps), external memory bandwidth (GBps),
thermal design power (Watts) and cost (\$), and for all models their compute and memory requirements;  
datapoints are shown in Sec.~\ref{sec:cmpmem} and \ref{sec:hw}. 
Combining application requirements with hardware platform characteristics can be leveraged for performance predictions using roofline models~\cite{rooflines}.  Level 0 is a base layer, with results that are available instantly, and provide a target point of reference, guidance for optimization efforts and allows to compute metrics such as achievable compute efficiency. At level 0, we already introduce the notion of performance per datatype operation which is essential to support quantization as an algorithmic optimization.

Two tables are presented in the appendix, one for hardware characteristics and one for neural networks. The hardware table has one row per hardware platform and supported native datatype; a minimum of Half Precision (FP16), Single Precision (FP32) and INT8 are recorded
\footnote{If INT8 is not natively supported, it can be embedded inside FP16}. 
In the second table, for each CNN, we record four values: total number of compute operations for a single input, the model size, the size of the state 
and the total amount of tensors in between layers that require buffering. These values can be used as a basis to derive memory requirements and compute requirements for both inference and training; examples are shown in Figure~\ref{fig:crt}.

{\bf Level 0 - Roofline Analysis}
Using assumptions for where weights, tensors, 
gradients, weight updates and state of a neural network are stored, combined with the size of the datatypes used, allow us to derive the arithmetic intensity of a neural network during training and inference. Combined with the roofline for a given hardware platform, we can provide insight as to whether a neural network will be memory or compute bound and guidance for what is theoretically possible (Fig.~\ref{fig:l0-0}) .

{\bf Level 1 - Compute Patterns}
Level 1 exposes achievable compute performance for typical compute patterns encountered within neural networks, which equates to popular layers including convolutions, fully connected layers, recurrent layers, residual layers, and squeeze layers, over a range of dimensions and with different numerical representations (Sec.~\ref{secNN}).  These tests are comparable to DeepBench~\cite{deepbench}, with the significant difference that we provide much broader support for specialized numerical representations.
For each of these compute patterns, and for both inference and training, we record the following figures of merit: measured performance (TOps or GOps), latency (ms),
power consumption (Watts) of the full platform in the embedded space, and of the board excluding the host system in the cloud.\footnote{Power measurements might not always  be available and might require specialized test infrastructures and testbeds.}
While level 1 does not capture application level accuracy, the tests will include verification of functional correctness.
The results should reflect achievable compute performance, excluding potential bottlenecks for moving data which are addressed in level 2. While requiring execution, the tests at level 1 are relatively rapid.  We include a sweep over batch and thread sizes.  

{\bf Level 2 - Compute \& Data Movement}
Level 2 
is comprised of simple combinations of level 1 tests, and can thereby effectively capture potential bottlenecks such as tensor movement between layers, as well as storage requirements.
It considers stacks of level 1 layers and only includes a subset of all possible combinations to keep test time to a minimum. 
We include mixed precision between layers in these small template stacks for both inference and training.
Figures of merit are identical to level 1. In particular, the latency variation between level 1 with single fused layers and level 2 with layer stacks will bring insight into data movement and buffering bottlenecks. 

{\bf Level 3 - Applications}
Application coverage is essential to offer space for algorithmic innovation which can achieve superior system-level performance and can only be validated when combined with application results. As such, achieved accuracy becomes the bar for normalizing results, and independent of the neural network. We include the initially planned datasets and models (Table~\ref{tableApps}), taken from existing benchmarks and complement these  with models that have been explored to work well with pruning and quantization optimizations. Furthermore, contributors are welcome to provide different models for given machine learning tasks. See Appendix for complete list. 

For inference, we include performance measurements for a single image. The error rate is the reported test error over the whole test dataset.
For training, we report throughput, training time (latency), and power for a single image as well (including correctness tests). 
We also provide measurements over longer training sequences with specific accuracy targets, for example, measure complete training time 90\% top5 error for ImageNet classification with a ResNet50. Finally, we offer the option to optimize the training algorithm and network and record all possible data points in a multi-dimensional graph; for those it is essential to include development time.
Similar concepts are being applied in MLPerf and Request \cite{mlperf, ckframework}.
There is no single criteria that decides whether one solution is optimal, as for different use cases, different figures of merit apply. All combinations yield different trade-offs within the multidimensional design space. As such, we present all solutions and measurements within multi-dimensional figures, whereby the pareto frontier represents the best possible compromises (Fig.~\ref{fig:pareto1}). 

{\bf Algorithmic Optimizations including Quantization}
This benchmarking proposal opens up the opportunity for algorithmic innovations. We include in this pruning and topological changes, while initially focusing on quantization and numerical representations.
For this, we include, on every level of the benchmark several numerical representations, including FP32, FP16, INT8, BIN, TERN, and allow for arbitrary choices to be included, for example Microsoft's custom floating point~\cite{microsoft-brainwave}. 
Training each neural network with different quantization approaches and different and potentially esoteric numerical representations is highly time-intensive..
Therefore, careful logging of trained quantized models is a high priority for level 3.

{\bf Frameworks \& Datasets}
Datasets are a key input to the benchmark and impact accuracy results.
We rely on open source datasets exclusively. Framework support is expected to be one of the biggest challenges since each framework is directly connected with a neural network and datasets within an application context and models are not necessarily portable. Therefore, 
we need operational hardware backends for a diverse set of AI accelerators which may or may not be available.
Furthermore, quantization is not necessarily mainstream in frameworks.
It is not yet clear to what extent cross compilation tools such as TVM \cite{tvm} can help, while exchange formats such as ONNX \cite{onnx} are still immature, lack adoption and very importantly full quantization support.  Training scripts exposing all hyperparameters, training initializations and so on must be fully logged as they can have significant impact on accuracy.

{\bf Power and Energy}
To represent power and energy cost, we only report platform power measured at the socket. While this is not necessarily accurate, there are strong reasons behind this choice.  
First, the measurement needs to be fair, therefore we believe subsystems, including memory specifically need to be taken into account.
Second, more detailed current sampling on the platforms may be available on some platforms, but each platform comes with different interfaces, and may or may not provide access to all power rails. 
While the accuracy of typical socket power meters is around 10\%, we found that these results remain representative of the systems. Furthermore, we average the results over 10 measurements. 

Another consideration is whether to consider power or energy per frame. We settled on using absolute power consumption since when multithreading or batching is applied, it is hard to derive a representative number for energy and would differ depending on whether the end application is latency or throughput driven. 
Finally, idle power with these platforms, can represent a significant percentage of the overall power budget and would therefore cloud the observation. In particular one FPGA platform is an evaluation board with many peripherals, which is reflected in high idle power (19.9 Watts) compared to the GPU (between 3.4 to 5.0 Watt depending on operating mode), while the additional dynamic power consumption is minimal and yields the FPGA overall as the more efficient platforms despite the initial load.

{\bf Testbeds, Reproducibility, \& Recorded Measurements}
In order to provide useful scientific results, all experiments and measurements must be validated and reproducible. Specifically: \\
$\bullet$ All input data to the test suites must be openly accessible. \\
$\bullet$ Many platforms can be made available through virtualized compute environments, which is adequate if the cost is not prohibitive. However some platforms may not be available. Therefore, an open testbed may be advisable and considered as an extension to this benchmark. \\
$\bullet$ As the higher levels of benchmarks may require a long time to run and hardware may not be available, we advocate recording of results, whereby each entry will be validated by a third party such that results are guaranteed to be a) reproducible and b) correct.

Our colleagues in the Request Tournament effort~\cite{request} leverage ACM's rigorous artifact evaluation technology and the Collective Knowledge Workflow Framework \cite{ckframework} and do an outstanding job addressing this. We aim to adopt the same principles.

{\bf Adaptability}
Machine Learning is currently a highly dynamic field, and specific algorithms may become very quickly outdated and new models may emerge and take over rapidly. We plan to adapt fast and add/retire models as machine learning science matures.

\section{Experimental Results \& Evaluation}
\label{sec:experiment}

We present measured results aimed at evaluating the defined benchmarking tests and figures of merit to ensure that they accurately reflect a system's capabilities. 
For test platforms, we used the Nvidia TX2 GPU and the Xilinx ZCU104 FPGA. For both platforms, we carried out all levels of tests on one specific Machine Learning task, ImageNet classification, for two different neural networks, GoogleNetV1 and ResNet50. We use FP32, FP16 (supported by GPU), and INT8 (supported by FPGA) as numerical representations, a form of algorithmic optimization. 
We run GPU platforms with a spectrum of batch sizes and different operating modes (MaxN, MaxQ, MaxP), which are optimized for different performance and power consumption targets\footnote{We also tested MaxP, however never achieved optimal values for any figure of merit.}. 
For FPGAs, there are a spectrum of implementations available. 
We exercise the Deephi DPU overlay, which uses threads instead of batch sizes to achieve high system utilization, and therefore exercise a spectrum of thread counts.
For FPGAs we show the theoretical limits of the current implementation (which is clocked at 666MHz), as well as the datasheet peak performance of 750MHz.
For GPUs, we use the theoretical peak as dictated by the clock frequencies defined by the operating mode.
Full experimental results are provided in the appendix. We currently have only exercised inference results to validate the benchmark methodology.
In the following, we evaluate each benchmarking level individually and then provide a first critical review of these early results. 

\begin{figure}
\centering
\includegraphics[width=\linewidth]{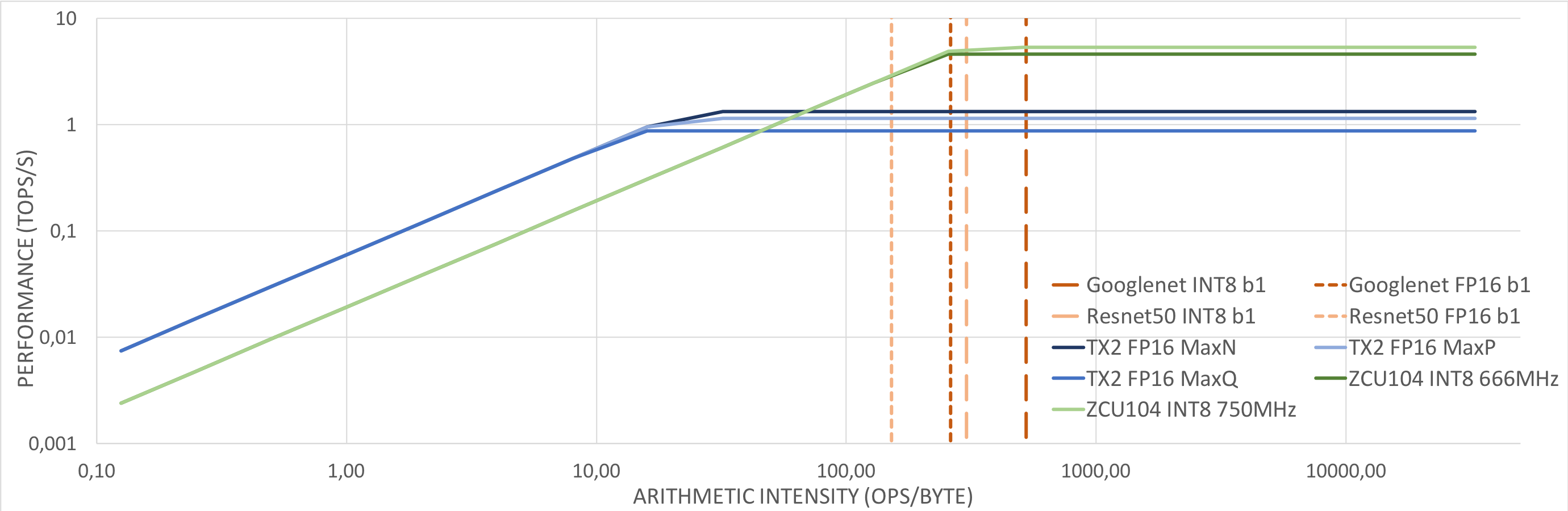}
\caption{TX2 and ZCU104 Level0 Rooflines with GoogleNet and ResNet50}
\label{fig:l0-0}
\end{figure}

{\bf Level 0} Using values for hardware platforms 
 and arithmetic intensity (AI) 
 we created rooflines for the target platforms and  performance predictions for both networks\footnote{We assumed that all weights are kept off-chip and all intermediate results are on-chip.  This assumption will be revised in the future.}.  Fig.~\ref{fig:l0-0} shows that both NNs 
 will be compute bound for INT8, FP16 and FP32. 
 The arithmetic intensity should be  higher for larger batch sizes (batch size of 1 is shown), but  the performance prediction for larger batch sizes will be identical.  The theoretical performance prediction can be derived from this and is summarized in table~\ref{l0-pp}. These numbers are  used to compute efficiency for levels 1, 2 and 3.

\small{
\begin{table}
\caption{Level 0 - Performance Predictions}
\label{l0-pp}
\resizebox{0.7\width}{!}{
\begin{tabular}{|l|cccc|cc|}
\toprule
\textbf{Performance} & \multicolumn{4}{c|}{\textbf{TX2}}& \multicolumn{2}{c|}{\textbf{ZCU104}}\\
\textbf{Predictions [TOPs]} & \textbf{FP32-MaxN} & \textbf{FP16-MaxN}&
\textbf{FP32-MaxQ} & \textbf{FP16-MaxQ}& \textbf{INT8-666MHz}& \textbf{INT8-750MHz}\\
\midrule
ResNet50 = GoogleNetV1   & 0.667 & 1.333 & 0.437 & 0.874 & 4.604 & 5.357 \\
\bottomrule
\end{tabular}
}
\end{table}}

{\bf Level~1 and Level~2}
We restrict the evaluation of level 1 and level 2 to ResNet50, as this is sufficient to make the key observations.
The ResNet50 topology is relatively regular in structure, consisting of a top convolutional layer with pooling combination, 16 residual blocks, and a fully connected layer. Each residual block is comprised of thresholding layers, convolutions, and elementwise additions.
As the convolutions account for the majority of the compute, we focus mainly on the convolutional layers of the network. Since the platform-specific frameworks perform layer fusion as network optimization, level 1 represents the smallest possible fused layer structure.
Table \ref{l1-var} shows level 1 and level 2 latency results for one TX2 hardware configuration (MaxN, FP16) with different batch sizes as well as level 1 results for ZCU104 with different thread numbers. We restrict level 1 to convolutions of different sizes and select the residual layers res2a, res3a, res4a and res5a to get an overview over the whole network. Level 2 results are provided for all residual layers of the network. Due to limited support by the hardware-specific framework, it is not possible to benchmark level 2 on FPGA platforms.
We observe a large discrepancy in execution time for different residual stacks, even though the compute requirements within each is similar. It is likely that data movement varies significantly depending on the incoming and outgoing tensor dimensions. Therefore, it is important to include as many layer types inside level 1 and 2 testing. We would expect this to be even more pronounced for other topologies, as they may be less balanced than ResNet50.
We also observe a large discrepancy between the performance of different convolutional layers (Table~\ref{l1-var}, level 1). Unlike the residual blocks, this is anticipated, as they come with very different compute requirements. Furthermore, the differences are more pronounced with larger batch size. It is therefore our plan to include the full spectrum of convolutional layers within level 1.

\begin{figure}[h]
\centering
\includegraphics[width=0.95\linewidth]{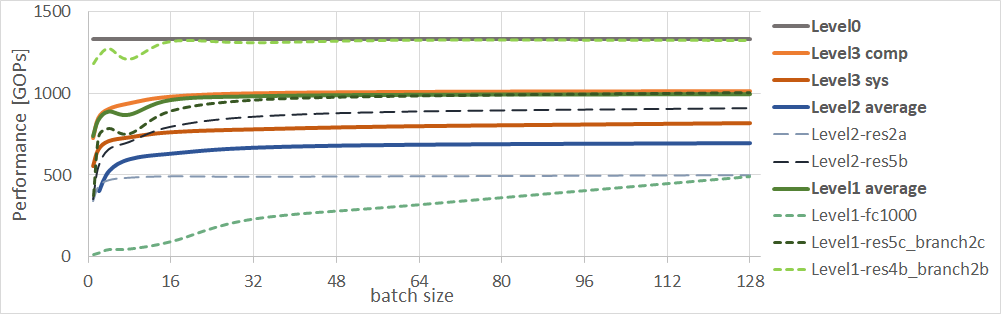}
\caption{Performance comparison layer0, layer1, layer2 and layer3 for TX2 (MaxN, FP16 configuration)}
\label{fig:l0-l3}
  \vspace*{-3ex}
\end{figure}

{\bf Multi-Tiered Concept}
Fig.~\ref{fig:l0-l3} depicts the performance measurements of the various levels. We restricted the visualized experiments to MaxN, FP16 configuration on TX2, and a subset of microbenchmarks on level 1 and level 2, for a spectrum of batch sizes. Note that the theoretical peak performance is significantly higher than measured performance, only within reach of individual layers that fit the hardware architecture well.
The system (level 3) achieves from 41.1 to 60.7\% efficiency, where larger batch sizes achieve higher performance. %This is in line with literature.
Level 2 results are on average more negative than achieved performance (level 3) and a fairly good approximation within 16\% of the achievable level 3 system performance, but far off level 3 compute performance. 
Level 1 results have usually better performance than the level 2 results. This makes intuitively sense, as a limited amount of bottlenecks are exposed during execution of the benchmark. In particular lower weight storage is required, which is most likely contained on-chip, thereby alleviating any potential memory bottlenecks. 
Also it can be said that the averaged level 1 results provide a good estimation of possible compute performance on level 3.
As already mentioned, for level 1 and 2 results, we observe large variations in performance ranges for different dimensions of convolutions. The insight is that to provide a good projection from level 1 or level 2 to level 3, we need to provide full coverage of convolutional layers.
Another challenge is that many backend tools perform automated layer fusion such as merging batch normalization with convolutions, which makes testing in isolation inaccurate.

{\bf Level 3 - Full system level performance evaluation}
The aim of level 3 is to explore optimal solutions within the design space regarding application performance 
independent of model topology and algorithmic optimizations. 
We include results for both platforms (TX2, ZCU104), for INT8, FP16, FP32, across the spectrum of batch sizes and thread numbers for both GoogleNetV1 and ResNet50.  
See plots of pareto points (Fig.~\ref{fig:l3-pareto}) and results in the Appendix.
We made the following key observations: Firstly, the ZCU104 FPGA provides the highest system level (948GOPs) and compute level performance (1067GOPs) compared to the GPU platform (809GOPs and 1011GOPs respectively) for both GoogleNetV1 and ResNet5050 (Fig.~\ref{fig:l3-pareto}, top left). For GoogleNetV1, the FPGA provides better performance and accuracy. For ResNet50, the FPGA provides better performance but lower accuracy compared to the GPU platform. Further, GoogleNetV1 topology provides more than 2x the performance compared to ResNet50, due to the significantly lower compute per frame required as part of the neural network topology, while ResNet50 provides best accuracy across the platforms. The accuracy difference is 1.59\% for the FPGA and 4.27\% for the GPU (Fig.~\ref{fig:l3-pareto} top left).
Additionally, the ZCU104 outperforms the TX2 in regards to latency by orders of magnitude and across topologies unless GPUs operate with small batch size, where the performance efficiency drops. GPU latency varies from a minimum of 8ms to a maximum of 1838.5ms for batch=128. FPGA latency varies from 9.65ms to 65ms.
Finally, the GPU platform is more power efficient,which can be attributed to the GPU platform being more optimized, whereas the FPGA platform is more general purpose. This is apparent when considering idle power (Sec.~\ref{sec:proposal}, 5 Watts for TX2 and 19.9 for ZCU104).

\begin{figure}[h]
\centering
\includegraphics[width=\linewidth]{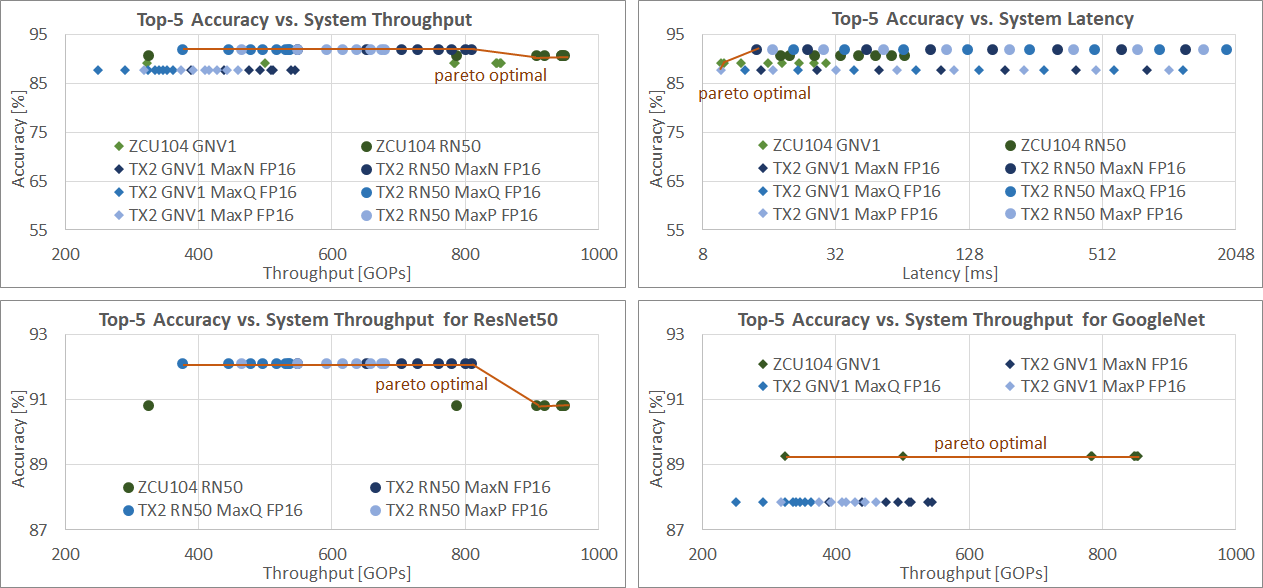}
\caption{Level 3: System Performance Evaluation}
\label{fig:l3-pareto}
  \vspace*{-3ex}
\end{figure}

In this evaluation above we consider full system-level performance (Fig.~\ref{fig:l3-pareto}), including initial data movement as well as compute only performance. Depending on the end application, it may be important to factor out the initial data movement from the overall time, as the inference engine might be included in a larger compute data path, where the inputs are streamed directly from on-chip resources. However when analyzing the experimental data points for both GPU and FPGA platforms, 
it appears that the difference is very regular in nature, and it is not obvious that a distinction within the benchmark is necessary (see Appendix) as long as it is clearly indicated what is measured. The pareto curves are an effective means to compare different topologies and different platforms leaving space for algorithmic optimizations. We plan to leverage 3- or 4-dimensional graphs to additionally explore 
relationships between latency and system-level performance.
 %Michaela
\section{Conclusion \& Future Work}
\label{sec:Con}

Neural networks are fast gaining popularity across an increasing number of applications. However, they are accompanied by challenging compute and memory requirements, as shown in Section~\ref{sec:cmpmem}, which is seriously challenging the semiconductor industry which is facing performance scalability issues.
This is of particular importance for embedded computing environments, where real estate, power and available compute and memory resources are at a premium.
As such the industry is turning to both algorithmic innovation in form of new topologies, quantization, and pruning strategies, as well as architectural innovation with more and more heterogeneous devices and the emergence of specialized DPUs. To facilitate better insights into the increasingly complex space of end solutions which involve hardware-software codesign and evaluate new concepts in computer architecture, novel NN system benchmarks are needed.

QuTiBench is a proposed novel benchmarking methodology to help drive hardware innovation and  provide insights for system level designers in understanding possible performance accuracy trade-offs for newly devised and fine-tuned algorithms combined with highly customized accelerators. 
Key contributions are that we provide concepts that allow benchmarking of highly optimized algorithms by tying hardware characteristics back to the end application, thereby providing the needed algorithmic freedom. Another key differentiator in this benchmarking concept is the introduction of the multi-tiered approach including a theoretical level and consideration of a spectrum of numerical representations at all levels. As such the benchmark can provide insights at various abstraction levels. This brings two key advantages: a) it provides a spectrum of insights and users can choose from instant but perhaps crude results, to elaborate results which require longer evaluation; and b) the multi-tiered approach provides insights into system bottlenecks. For example, are the 
recurrent or the fully connected layers the challenge? Or is the bottleneck the data movement in between?  
We present initial experimental results on two types of neural network topologies aimed at image classification tasks, and exercise them on two different types of hardware platforms for all levels of the proposed benchmarks. We present some of the lessons learned while exercising the benchmarks, challenges encountered, and analyze the quality of the results in regards to real system performance at the various levels.

This effort is just beginning. Future work will focus on refining details and running broader experimentation. We plan to expand on level 0 results first and build out test suites targeting FPGAs, GPUs, CPUs and DPUs within the embedded space. Many concepts regarding reproducibility need to be refined, as well as automated software testing infrastructure as proposed by deep500.org. Also collaboration with larger efforts such as MLPerf will be beneficial to gain traction. We invite the research community to contribute to QuTiBench.  
  %Michaela

\begin{acks}
The authors would like to thank the FINN team at Xilinx research, Prof.~Ce Zhang at ETH Zurich, and the Deephi team for insights and support.  Miriam Leeser is supported in part by the National Science Foundation under Grant No. 1717213.
\end{acks}

% Bibliography
\bibliographystyle{ACM-Reference-Format}
\bibliography{bibliography}

\clearpage
% Appendix
\appendix

\section{Appendix: Table of Results}
\label{sec:appendix}

\begin{table}[!ht]
\caption{Planned Applications, Datasets and Models}
\label{planned}
\resizebox{\textwidth}{!}{
\begin{tabular}{|lll|cc|}
\toprule
\textbf{Learning Technique} & \textbf{Application} &  & \multicolumn{2}{c|}{\textbf{QuTiBench}}\\
 & & & \textbf{ Dataset} & \textbf{Model}\\\midrule
\textbf{Supervised} & \textbf{Vision} & \textbf{Image Classification} 
& ImageNet, MNIST & ResNet50, MobileNet(V1), GoogleNet, MLP \\
& \textbf{Vision} & \textbf{Object Detection} 
& Pascal VOC & SSD-ResNet34, YoloV2\\
& \textbf{Vision} & \textbf{Semantic Segmentation} 
&  Pascal VOC & Mask R-CNN, SSD-MobileNet\\
& \textbf{NLP} & \textbf{Machine Translation}  
& WMT'14 English-to-French\&German & GNMT~\cite{wu2016google} \\
& \textbf{NLP} & \textbf{Speech Recognition} 
& Librispeech & DeepSpeech2\\
& \textbf{NLP} & \textbf{Sentiment Analysis}  
& SST, IMDB, SemEval2018 & Multiplicative LSTM\\
& \textbf{NLP} & \textbf{Language Modeling}  
& babI & Memory Network \\
& \textbf{Recommendation} & \textbf{Movies}
& Movielens 20M & NCF\\
\midrule
\textbf{Unsupervised} & \textbf{Vision} & \textbf{Feature Extraction}  
& MNIST & autoencoder\\
 & \textbf{Vision} & \textbf{Adversarial Learning}  
& ImangeNet & WGAN \\
\midrule
\textbf{Deep Reinforcement Learning} & \textbf{Game} & \textbf{Go}  
& Go & MiniGo \\
& & \textbf{Atari ALE}  & Atari ALE & DeepQ \\
\bottomrule
\end{tabular}
}
\end{table}

\begin{table}[!ht]
\caption{Level 0 - Hardware Platforms \& Neural Network Model}
\label{tab:level0-xs}
\resizebox{\linewidth}{!}{
\begin{tabular}{|l|l||cccc||l|ccc|}
\toprule
\textbf{Hardware Platform} & \textbf{datatype} & \multicolumn{4}{c||}{\textbf{Figures of Merit (theo.)}} &
\textbf{Model} & \multicolumn{3}{c|}{\textbf{Figures of Merit (theo.)}}\\
&& \textbf{[TOPs]} & \textbf{[GBps]} & \textbf{[Watts]} & 
\textbf{[\$]} & & \textbf{[GOP]} & \textbf{Size [ME]} & \textbf{AI [OP:Byte]}\\
\midrule
Nvidia Jetson TX2 MaxN & FP32 & 0.67 & 59.7 & NA & 469 &
ResNet50 (b=1, INT8) & 7.72 & 25.50 & 303 \\
Nvidia Jetson TX2 MaxP & FP32 & 0.57 & 59.7 & 15.0 & 469 &
ResNet50 (b=8, INT8)& 7.72 & 25.50 & 2422 \\
Nvidia Jetson TX2 MaxQ & FP32 & 0.44 & 59.7 & 7.5 & 469 &
ResNet50 (b=1, FP16) & 7.72 & 25.50 & 151 \\
Nvidia Jetson TX2 MaxN & FP16 & 1.33 & 59.7 & NA & 469 &
ResNet50 (b=8, FP16) & 7.72 & 25.50 & 1211 \\
Nvidia Jetson TX2 MaxP & FP16 & 1.15 & 59.7 & 15.0 & 469 &
GoogleNetV1 (b=1, INT8) & 3.13 & 5.98 & 523\\
Nvidia Jetson TX2 MaxQ & FP16 & 0.87 & 59.7 & 7.5 & 469 &
GoogleNetV1 (b=8, INT8) & 3.13 & 5.98 & 4188\\
Xilinx ZCU104 DPU 666MHz & INT8 & 4.60 & 19.2 & NA & 895 &
GoogleNetV1 (b=1, FP16)& 3.13 & 5.98 & 262\\
Xilinx ZCU104 DPU 775MHz & INT8 & 5.36 & 19.2 & NA & 895 &
GoogleNetV1 (b=8, FP16) & 3.13 & 5.98 & 2094\\
\bottomrule
\end{tabular}
}
\end{table}

\begin{table}[!ht]
	\caption{Level 1 -ZCU104 Inference Results ResNet50 Individual Convolutional Layers}
	\label{AT_L1_FPGA}
	\resizebox{\linewidth}{!}{
		\begin{tabular}{|l||cccccc||cc|cc|}
			\toprule
			 \textbf{ZCU104} & \multicolumn{6}{c||}{\textbf{Network Parameters}} & \multicolumn{4}{c|}{\textbf{Figures of Merit}} \\ 
			& & & & & & & \multicolumn{2}{c|}{\textbf{thread=1}} & \multicolumn{2}{c|}{\textbf{thread=8}} \\
			\textbf{Layer} & 
			\textbf{[MOP]} & \textbf{in dim} &\textbf{in ch} & \textbf{filter} & \textbf{stride} & \textbf{out ch} &			
			\textbf{Latency} & \textbf{Throughput (Eff)} & \textbf{Latency} & \textbf{Throughput (Eff)}\\
			& & & & & & & \textbf{[ms]} & \textbf{[GOPs]} & \textbf{[ms]} & \textbf{[GOPs]}\\
			\midrule
res2a\_branch2a & 25.7 & 56 & 64 & 1 & 1 & 64 & 0.060 & 428.05 (0.09) & 0.082 & 630.21 (0.14)\\
res2a\_branch2b & 231.2 & 56 & 64 & 3 & 1 & 64 & 0.190 & 1216.84 (0.26) & 0.190 & 2428.70 (0.53)\\
res2a\_branch2c & 102.8 & 56 & 64 & 1 & 1 & 256 & 0.220 & 467.23 (0.10) & 0.258 & 798.45 (0.17)\\
res2a\_branch1 & 102.8 & 56 & 64 & 1 & 1 & 256 & 0.430 & 239.08 (0.05) & 0.464 & 443.34 (0.10)\\
res2b\_branch2a & 102.8 & 56 & 256 & 1 & 1 & 64 & 0.142 & 725.63 (0.16) & 0.196 & 1049.57 (0.23)\\
res2b\_branch2b & 231.2 & 56 & 64 & 3 & 1 & 64 & 0.190 & 1216.84 (0.26) & 0.190 & 2428.19 (0.53)\\
res2b\_branch2c & 102.8 & 56 & 64 & 1 & 1 & 256 & 0.429 & 239.64 (0.05) & 0.463 & 443.98 (0.10)\\
res2c\_branch2a & 102.8 & 56 & 256 & 1 & 1 & 64 & 0.140 & 734.13 (0.16) & 0.193 & 1063.14 (0.23)\\
res2c\_branch2b & 231.2 & 56 & 64 & 3 & 1 & 64 & 0.190 & 1216.84 (0.26) & 0.190 & 2428.32 (0.53)\\
res2c\_branch2c & 102.8 & 56 & 64 & 1 & 1 & 256 & 0.435 & 236.20 (0.05) & 0.462 & 444.69 (0.10)\\
res3a\_branch2a & 51.4 & 28 & 256 & 1 & 2 & 128 & 0.090 & 571.05 (0.12) & 0.128 & 800.12 (0.17)\\
res3a\_branch2b & 231.2 & 28 & 128 & 3 & 1 & 128 & 0.210 & 1100.95 (0.24) & 0.214 & 2159.44 (0.47)\\
res3a\_branch2c & 102.8 & 28 & 128 & 1 & 1 & 512 & 0.210 & 489.52 (0.11) & 0.247 & 832.79 (0.18)\\
res3a\_branch1 & 205.5 & 28 & 256 & 1 & 2 & 512 & 0.330 & 622.71 (0.14) & 0.390 & 1052.87 (0.23)\\
res3b\_branch2a & 102.8 & 28 & 512 & 1 & 1 & 128 & 0.120 & 856.45 (0.19) & 0.148 & 1391.07 (0.30)\\
res3b\_branch2b & 231.2 & 28 & 128 & 3 & 1 & 128 & 0.210 & 1100.95 (0.24) & 0.214 & 2165.20 (0.47)\\
res3b\_branch2c & 102.8 & 28 & 128 & 1 & 1 & 512 & 0.320 & 321.24 (0.07) & 0.353 & 582.88 (0.13)\\
res3c\_branch2a & 102.8 & 28 & 512 & 1 & 1 & 128 & 0.120 & 856.60 (0.19) & 0.151 & 1361.41 (0.30)\\
res3c\_branch2b & 231.2 & 28 & 128 & 3 & 1 & 128 & 0.210 & 1100.95 (0.24) & 0.215 & 2154.81 (0.47)\\
res3c\_branch2c & 102.8 & 28 & 128 & 1 & 1 & 512 & 0.303 & 339.02 (0.07) & 0.354 & 580.14 (0.13)\\
res3d\_branch2a & 102.8 & 28 & 512 & 1 & 1 & 128 & 0.120 & 856.52 (0.19) & 0.149 & 1383.86 (0.30)\\
res3d\_branch2b & 231.2 & 28 & 128 & 3 & 1 & 128 & 0.210 & 1100.95 (0.24) & 0.214 & 2165.50 (0.47)\\
res3d\_branch2c & 102.8 & 28 & 128 & 1 & 1 & 512 & 0.301 & 341.20 (0.07) & 0.353 & 582.72 (0.13)\\
res4a\_branch2a & 51.4 & 14 & 512 & 1 & 2 & 256 & 0.120 & 428.80 (0.09) & 0.133 & 774.21 (0.17)\\
res4a\_branch2b & 231.2 & 14 & 256 & 3 & 1 & 256 & 0.210 & 1100.95 (0.24) & 0.230 & 2011.48 (0.44)\\
res4a\_branch2c & 102.8 & 14 & 256 & 1 & 1 & 1024 & 0.290 & 354.46 (0.08) & 0.379 & 541.92 (0.12)\\
res4a\_branch1 & 205.5 & 14 & 512 & 1 & 2 & 1024 & 0.430 & 477.87 (0.10) & 0.500 & 821.34 (0.18)\\
res4b\_branch2a & 102.8 & 14 & 1024 & 1 & 1 & 256 & 0.130 & 790.71 (0.17) & 0.162 & 1271.41 (0.28)\\
res4b\_branch2b & 231.2 & 14 & 256 & 3 & 1 & 256 & 0.210 & 1100.95 (0.24) & 0.229 & 2015.61 (0.44)\\
res4b\_branch2c & 102.8 & 14 & 256 & 1 & 1 & 1024 & 0.350 & 293.69 (0.06) & 0.436 & 471.20 (0.10)\\
res4c\_branch2a & 102.8 & 14 & 1024 & 1 & 1 & 256 & 0.130 & 790.71 (0.17) & 0.163 & 1263.60 (0.27)\\
res4c\_branch2b & 231.2 & 14 & 256 & 3 & 1 & 256 & 0.210 & 1100.95 (0.24) & 0.231 & 2002.86 (0.44)\\
res4c\_branch2c & 102.8 & 14 & 256 & 1 & 1 & 1024 & 0.360 & 285.52 (0.06) & 0.438 & 469.22 (0.10)\\
res4d\_branch2a & 102.8 & 14 & 1024 & 1 & 1 & 256 & 0.130 & 790.65 (0.17) & 0.164 & 1251.14 (0.27)\\
res4d\_branch2b & 231.2 & 14 & 256 & 3 & 1 & 256 & 0.210 & 1100.95 (0.24) & 0.229 & 2019.57 (0.44)\\
res4d\_branch2c & 102.8 & 14 & 256 & 1 & 1 & 1024 & 0.350 & 293.76 (0.06) & 0.425 & 484.02 (0.11)\\
res4e\_branch2a & 102.8 & 14 & 1024 & 1 & 1 & 256 & 0.130 & 790.71 (0.17) & 0.162 & 1267.18 (0.28)\\
res4e\_branch2b & 231.2 & 14 & 256 & 3 & 1 & 256 & 0.210 & 1100.90 (0.24) & 0.230 & 2014.73 (0.44)\\
res4e\_branch2c & 102.8 & 14 & 256 & 1 & 1 & 1024 & 0.350 & 293.68 (0.06) & 0.438 & 469.19 (0.10)\\
res4f\_branch2a & 102.8 & 14 & 1024 & 1 & 1 & 256 & 0.130 & 790.53 (0.17) & 0.162 & 1265.78 (0.27)\\
res4f\_branch2b & 231.2 & 14 & 256 & 3 & 1 & 256 & 0.210 & 1100.95 (0.24) & 0.230 & 2007.12 (0.44)\\
res4f\_branch2c & 102.8 & 14 & 256 & 1 & 1 & 1024 & 0.360 & 285.49 (0.06) & 0.421 & 488.23 (0.11)\\
res5a\_branch2a & 51.4 & 7 & 1024 & 1 & 2 & 512 & 0.120 & 427.94 (0.09) & 0.188 & 546.52 (0.12)\\
res5a\_branch2b & 231.2 & 7 & 512 & 3 & 1 & 512 & 0.330 & 699.93 (0.15) & 0.493 & 937.72 (0.20)\\
res5a\_branch2c & 102.8 & 7 & 512 & 1 & 1 & 2048 & 0.470 & 218.66 (0.05) & 0.600 & 342.79 (0.07)\\
res5a\_branch1 & 205.5 & 7 & 1024 & 1 & 2 & 2048 & 0.517 & 397.60 (0.09) & 0.691 & 594.55 (0.13)\\
res5b\_branch2a & 102.8 & 7 & 2048 & 1 & 1 & 512 & 0.170 & 604.28 (0.13) & 0.272 & 755.16 (0.16)\\
res5b\_branch2b & 231.2 & 7 & 512 & 3 & 1 & 512 & 0.331 & 698.07 (0.15) & 0.499 & 926.34 (0.20)\\
res5b\_branch2c & 102.8 & 7 & 512 & 1 & 1 & 2048 & 0.500 & 205.56 (0.04) & 0.628 & 327.34 (0.07)\\
res5c\_branch2a & 102.8 & 7 & 2048 & 1 & 1 & 512 & 0.170 & 604.49 (0.13) & 0.265 & 775.00 (0.17)\\
res5c\_branch2b & 231.2 & 7 & 512 & 3 & 1 & 512 & 0.340 & 679.68 (0.15) & 0.503 & 918.94 (0.20)\\
res5c\_branch2c & 102.8 & 7 & 512 & 1 & 1 & 2048 & 0.500 & 205.55 (0.04) & 0.632 & 325.22 (0.07)\\
			\midrule
			\midrule
		\end{tabular}
	}
	\vspace*{-3ex}
\end{table}

\begin{table}[!ht]
	\caption{Level 1 -TX2 (MaxN, FP16) Inference Results ResNet50 Individual Convolutional Layers}
	\label{AT_L1_GPU}
	\resizebox{\linewidth}{!}{
		\begin{tabular}{|l||cccccc||cc|cc|}
			\toprule
			 \textbf{TX2} & \multicolumn{6}{c||}{\textbf{Network Parameters}} & \multicolumn{4}{c|}{\textbf{Figures of Merit}} \\ 
			& & & & & & & \multicolumn{2}{c|}{\textbf{MaxN, FP16, batch=1}} & \multicolumn{2}{c|}{\textbf{MaxN, FP16, batch=128}} \\
			\textbf{Layer} & 
			\textbf{[MOP]} & \textbf{in dim} &\textbf{in ch} & \textbf{filter} & \textbf{stride} & \textbf{out ch} &			
			\textbf{Latency} & \textbf{Throughput (Eff)} & \textbf{Latency} & \textbf{Throughput (Eff)}\\
			& & & & & & & \textbf{[ms]} & \textbf{[GOPs]} & \textbf{[ms]} & \textbf{[GOPs]}\\
			\midrule
res2a\_branch2a & 25.7 & 56 & 64 & 1 & 1 & 64 & 0.06 & 414.52 (0.31) & 5.05 & 651.15 (0.49)\\
res2a\_branch2b & 231.2 & 56 & 64 & 3 & 1 & 64 & 0.19 & 1197.93 (0.90) & 22.78 & 1299.39 (0.97)\\
res2a\_branch2c & 102.8 & 56 & 64 & 1 & 1 & 256 & 0.18 & 577.53 (0.43) & 20.15 & 653.15 (0.49)\\
res2a\_branch1 & 102.8 & 56 & 64 & 1 & 1 & 256 & 0.21 & 487.20 (0.37) & 23.66 & 556.19 (0.42)\\
res2b\_branch2a & 102.8 & 56 & 256 & 1 & 1 & 64 & 0.13 & 778.79 (0.58) & 13.74 & 957.60 (0.72)\\
res2b\_branch2b & 231.2 & 56 & 64 & 3 & 1 & 64 & 0.19 & 1210.47 (0.91) & 22.87 & 1293.82 (0.97)\\
res2b\_branch2c & 102.8 & 56 & 64 & 1 & 1 & 256 & 0.21 & 489.52 (0.37) & 23.68 & 555.77 (0.42)\\
res2c\_branch2a & 102.8 & 56 & 256 & 1 & 1 & 64 & 0.13 & 784.73 (0.59) & 13.74 & 957.88 (0.72)\\
res2c\_branch2b & 231.2 & 56 & 64 & 3 & 1 & 64 & 0.19 & 1210.47 (0.91) & 22.85 & 1295.01 (0.97)\\
res2c\_branch2c & 102.8 & 56 & 64 & 1 & 1 & 256 & 0.21 & 489.52 (0.37) & 23.67 & 555.82 (0.42)\\
res3a\_branch2a & 51.4 & 28 & 256 & 1 & 2 & 128 & 0.09 & 584.09 (0.44) & 7.19 & 915.30 (0.69)\\
res3a\_branch2b & 231.2 & 28 & 128 & 3 & 1 & 128 & 0.21 & 1095.73 (0.82) & 24.63 & 1201.33 (0.90)\\
res3a\_branch2c & 102.8 & 28 & 128 & 1 & 1 & 512 & 0.15 & 694.59 (0.52) & 15.18 & 866.82 (0.65)\\
res3a\_branch1 & 205.5 & 28 & 256 & 1 & 2 & 512 & 0.29 & 718.53 (0.54) & 30.35 & 866.60 (0.65)\\
res3b\_branch2a & 102.8 & 28 & 512 & 1 & 1 & 128 & 0.13 & 767.16 (0.58) & 12.07 & 1090.45 (0.82)\\
res3b\_branch2b & 231.2 & 28 & 128 & 3 & 1 & 128 & 0.21 & 1106.22 (0.83) & 24.66 & 1199.92 (0.90)\\
res3b\_branch2c & 102.8 & 28 & 128 & 1 & 1 & 512 & 0.16 & 634.57 (0.48) & 16.68 & 788.87 (0.59)\\
res3c\_branch2a & 102.8 & 28 & 512 & 1 & 1 & 128 &0.13 & 767.16 (0.58) & 12.10 & 1087.11 (0.82)\\
res3c\_branch2b & 231.2 & 28 & 128 & 3 & 1 & 128 & 0.21 & 1100.95 (0.83) & 24.47 & 1209.19 (0.91)\\
res3c\_branch2c & 102.8 & 28 & 128 & 1 & 1 & 512 & 0.16 & 634.57 (0.48) & 16.71 & 787.65 (0.59)\\
res3d\_branch2a & 102.8 & 28 & 512 & 1 & 1 & 128 & 0.13 & 767.16 (0.58) & 12.12 & 1085.95 (0.81)\\
res3d\_branch2b & 231.2 & 28 & 128 & 3 & 1 & 128 & 0.21 & 1106.22 (0.83) & 24.69 & 1198.56 (0.90)\\
res3d\_branch2c & 102.8 & 28 & 128 & 1 & 1 & 512 & 0.16 & 630.67 (0.47) & 16.67 & 789.35 (0.59)\\
res4a\_branch2a & 51.4 & 14 & 512 & 1 & 2 & 256 & 0.08 & 642.50 (0.48) & 7.10 & 926.26 (0.69)\\
res4a\_branch2b & 231.2 & 14 & 256 & 3 & 1 & 256 & 0.20 & 1185.64 (0.89) & 23.12 & 1279.89 (0.96)\\
res4a\_branch2c & 102.8 & 14 & 256 & 1 & 1 & 1024 & 0.15 & 708.97 (0.53) & 13.01 & 1011.64 (0.76)\\
res4a\_branch1 & 205.5 & 14 & 512 & 1 & 2 & 1024 & 0.28 & 728.72 (0.55) & 29.23 & 899.87 (0.68)\\
res4b\_branch2a & 102.8 & 14 & 1024 & 1 & 1 & 256 & 0.13 & 784.73 (0.59) & 11.55 & 1139.45 (0.85)\\
res4b\_branch2b & 231.2 & 14 & 256 & 3 & 1 & 256 & 0.20 & 1179.59 (0.88) & 22.33 & 1325.28 (0.99)\\
res4b\_branch2c & 102.8 & 14 & 256 & 1 & 1 & 1024 & 0.15 & 680.79 (0.51) & 13.75 & 957.18 (0.72)\\
res4c\_branch2a & 102.8 & 14 & 1024 & 1 & 1 & 256 & 0.13 & 778.79 (0.58) & 11.62 & 1132.10 (0.85)\\
res4c\_branch2b & 231.2 & 14 & 256 & 3 & 1 & 256 & 0.20 & 1173.60 (0.88) & 22.99 & 1287.35 (0.97)\\
res4c\_branch2c & 102.8 & 14 & 256 & 1 & 1 & 1024 & 0.15 & 680.79 (0.51) & 13.76 & 956.14 (0.72)\\
res4d\_branch2a & 102.8 & 14 & 1024 & 1 & 1 & 256 & 0.13 & 778.79 (0.58) & 11.57 & 1137.09 (0.85)\\
res4d\_branch2b & 231.2 & 14 & 256 & 3 & 1 & 256 & 0.20 & 1185.64 (0.89) & 22.92 & 1291.17 (0.97)\\
res4d\_branch2c & 102.8 & 14 & 256 & 1 & 1 & 1024 & 0.15 & 680.79 (0.51) & 13.76 & 956.00 (0.72)\\
res4e\_branch2a & 102.8 & 14 & 1024 & 1 & 1 & 256 & 0.13 & 778.79 (0.58) & 11.59 & 1135.32 (0.85)\\
res4e\_branch2b & 231.2 & 14 & 256 & 3 & 1 & 256 & 0.20 & 1185.64 (0.89) & 22.85 & 1295.41 (0.97)\\
res4e\_branch2c & 102.8 & 14 & 256 & 1 & 1 & 1024 & 0.15 & 680.79 (0.51) & 13.78 & 954.89 (0.72)\\
res4f\_branch2a & 102.8 & 14 & 1024 & 1 & 1 & 256 & 0.13 & 784.73 (0.59) & 11.65 & 1129.96 (0.85)\\
res4f\_branch2b & 231.2 & 14 & 256 & 3 & 1 & 256 & 0.20 & 1179.59 (0.88) & 22.40 & 1321.26 (0.99)\\
res4f\_branch2c & 102.8 & 14 & 256 & 1 & 1 & 1024 & 0.15 & 680.79 (0.51) & 13.78 & 955.17 (0.72)\\
res5a\_branch2a & 51.4 & 7 & 1024 & 1 & 2 & 512 & 0.14 & 372.46 (0.28) & 7.61 & 864.77 (0.65)\\
res5a\_branch2b & 231.2 & 7 & 512 & 3 & 1 & 512 & 0.31 & 748.22 (0.56) & 24.90 & 1188.59 (0.89)\\
res5a\_branch2c & 102.8 & 7 & 512 & 1 & 1 & 2048 & 0.27 & 386.47 (0.29) & 12.53 & 1049.90 (0.79)\\
res5a\_branch1 & 205.5 & 7 & 1024 & 1 & 2 & 2048 & 0.51 & 406.93 (0.31) & 30.92 & 850.85 (0.64)\\
res5b\_branch2a & 102.8 & 7 & 2048 & 1 & 1 & 512 & 0.22 & 475.93 (0.36) & 11.35 & 1159.74 (0.87)\\
res5b\_branch2b & 231.2 & 7 & 512 & 3 & 1 & 512 & 0.30 & 763.04 (0.57) & 24.91 & 1188.26 (0.89)\\
res5b\_branch2c & 102.8 & 7 & 512 & 1 & 1 & 2048 & 0.27 & 382.16 (0.29) & 13.21 & 995.87 (0.75)\\
res5c\_branch2a & 102.8 & 7 & 2048 & 1 & 1 & 512 & 0.22 & 473.73 (0.36) & 11.39 & 1155.36 (0.87)\\
res5c\_branch2b & 231.2 & 7 & 512 & 3 & 1 & 512 & 0.31 & 753.09 (0.56) & 24.91 & 1187.88 (0.89)\\
res5c\_branch2c & 102.8 & 7 & 512 & 1 & 1 & 2048 & 0.28 & 371.12 (0.28) & 13.09 & 1005.53 (0.75)\\
			\midrule
			\midrule
		\end{tabular}
	}
	\vspace*{-3ex}
\end{table}

\begin{table}[!ht]
	\caption{Level 2 - Inference Results ResNet50 Residual Layers}
	\label{AT_L2}
	\resizebox{0.7\width}{!}{
		\begin{tabular}{|lll||cc|cc|cc|}
			\toprule
			 & & &  \multicolumn{2}{c|}{\textbf{MaxN}} & \multicolumn{2}{c|}{\textbf{MaxQ}}  & \multicolumn{2}{c|}{\textbf{MaxP}}\\
			\textbf{HW} & \textbf{Layer} & \textbf{Parameters} & \textbf{Lat} &  \textbf{Throughput (Eff)} & \textbf{Lat} & \textbf{Throughput (Eff)} & \textbf{Lat} &  \textbf{Throughput (Eff)}\\
			 & & & \textbf{[ms]} & \textbf{[GOPs] ([\%])} & \textbf{[ms]} & \textbf{[GOPs] ([\%])} & \textbf{[ms]} & \textbf{[GOPs] ([\%])}  \\
			\midrule
    TX2 &  res2a & FP16, b=1 & 1.37 & 431.27 (0.32) & 1.90 & 292.13 (0.25) & 1.58 & 371.40 (0.42)\\
    TX2 &  res2a & FP16, b=2 & 2.17 & 464.25 (0.35) & 3.08 & 314.23 (0.27) & 2.50 & 401.64 (0.46)\\
    TX2 &  res2a & FP16, b=4 & 3.97 & 481.73 (0.36) & 5.83 & 325.90 (0.28) & 4.61 & 418.69 (0.48)\\
    TX2 &  res2a & FP16, b=8 & 7.69 & 491.73 (0.37) & 11.30 & 330.22 (0.29) & 8.91 & 426.23 (0.49)\\
    TX2 &  res2a & FP16, b=16 & 15.11 & 495.85 (0.37) & 22.39 & 333.24 (0.29) & 17.48 & 428.54 (0.49)\\
    TX2 &  res2a & FP16, b=32 & 30.39 & 436.04 (0.33) & 44.20 & 333.95 (0.29) & 34.49 & 430.49 (0.49)\\
    TX2 &  res2a & FP16, b=64 & 60.41 & 492.24 (0.37) & 88.98 & 334.18 (0.29) & 68.93 & 430.10 (0.49)\\
    TX2 &  res2a & FP16, b=128 & 119.12 & 495.33 (0.37) & 177.37 & 333.95 (0.29) & 137.24 & 430.10 (0.49)\\ \midrule
    TX2 &  res2b & FP16, b=1 & 1.12 & 443.23 (0.33) & 1.57 & 303.00 (0.26) & 1.32 & 382.90 (0.44)\\
    TX2 &  res2b & FP16, b=2 & 1.95 & 481.92 (0.36) & 2.77 & 333.25 (0.29) & 2.29 & 416.41 (0.48)\\
    TX2 &  res2b & FP16, b=4 & 3.63 & 502.50 (0.38) & 5.24 & 343.22 (0.30) & 4.19 & 437.17 (0.50)\\
    TX2 &  res2b & FP16, b=8 & 6.95 & 509.95 (0.38) & 10.06 & 352.14 (0.31) & 7.99 & 445.00 (0.51)\\
    TX2 &  res2b & FP16, b=16 & 13.62 & 515.82 (0.39) & 19.90 & 354.66 (0.31) & 15.86 & 448.12 (0.51)\\
    TX2 &  res2b & FP16, b=32 & 27.19 & 518.82 (0.39) & 39.57 & 356.64 (0.31) & 31.05 & 451.28 (0.52)\\
    TX2 &  res2b & FP16, b=64 & 54.32 & 515.82 (0.39) & 78.80 & 356.35 (0.31) & 62.09 & 452.20 (0.52)\\
    TX2 &  res2b & FP16, b=128 & 108.24 & 517.02 (0.39) & 158.53 & 355.50 (0.31) & 124.29 & 450.38 (0.52)\\ \midrule
    TX2 &  res2c & FP16, b=1 & 1.12 & 446.33 (0.33) & 1.59 & 301.77 (0.26) & 1.32 & 380.94 (0.44)\\
    TX2 &  res2c & FP16, b=2 & 1.96 & 483.48 (0.36) & 2.75 & 333.75 (0.29) & 2.27 & 419.53 (0.48)\\
    TX2 &  res2c & FP16, b=4 & 3.60 & 504.19 (0.38) & 5.16 & 346.15 (0.30) & 4.17 & 438.88 (0.50)\\
    TX2 &  res2c & FP16, b=8 & 6.89 & 512.87 (0.38) & 10.12 & 347.49 (0.30) & 8.04 & 446.33 (0.51)\\
    TX2 &  res2c & FP16, b=16 & 13.59 & 516.42 (0.39) & 19.86 & 355.50 (0.31) & 15.70 & 451.28 (0.52)\\
    TX2 &  res2c & FP16, b=32 & 27.01 & 518.22 (0.39) & 39.22 & 356.92 (0.31) & 31.14 & 451.74 (0.52)\\
    TX2 &  res2c & FP16, b=64 & 54.36 & 515.82 (0.39) & 78.66 & 356.64 (0.31) & 62.06 & 448.57 (0.51)\\
    TX2 &  res2c & FP16, b=128 & 108.07 & 516.42 (0.39) & 158.11 & 355.22 (0.31) & 123.99 & 449.92 (0.51)\\ \midrule
    TX2 &  res3a & FP16, b=1 & 1.39 & 475.68 (0.36) & 1.96 & 323.22 (0.28) & 1.66 & 406.90 (0.47)\\
    TX2 &  res3a & FP16, b=2 & 2.38 & 523.41 (0.39) & 3.45 & 356.13 (0.31) & 2.81 & 449.19 (0.51)\\
    TX2 &  res3a & FP16, b=4 & 4.39 & 555.61 (0.42) & 6.43 & 374.19 (0.33) & 5.18 & 476.43 (0.55)\\
    TX2 &  res3a & FP16, b=8 & 8.46 & 563.90 (0.42) & 12.39 & 385.63 (0.34) & 9.80 & 490.72 (0.56)\\
    TX2 &  res3a & FP16, b=16 & 16.53 & 575.15 (0.43) & 24.45 & 390.61 (0.34) & 19.22 & 495.95 (0.57)\\
    TX2 &  res3a & FP16, b=32 & 32.86 & 576.80 (0.43) & 48.62 & 391.37 (0.34) & 38.07 & 498.40 (0.57)\\
    TX2 &  res3a & FP16, b=64 & 65.46 & 577.35 (0.43) & 96.24 & 393.92 (0.34) & 75.66 & 500.05 (0.57)\\
    TX2 &  res3a & FP16, b=128 & 133.09 & 568.13 (0.43) & 194.49 & 389.61 (0.34) & 151.72 & 496.77 (0.57)\\ \midrule
    TX2 &  res3b & FP16, b=1 & 1.03 & 511.11 (0.38) & 1.41 & 347.49 (0.30) & 1.21 & 438.45 (0.50)\\
    TX2 &  res3b & FP16, b=2 & 1.69 & 562.54 (0.42) & 2.40 & 385.54 (0.34) & 1.96 & 488.23 (0.56)\\
    TX2 &  res3b & FP16, b=4 & 3.05 & 597.89 (0.45) & 4.43 & 407.31 (0.35) & 3.58 & 517.02 (0.59)\\
    TX2 &  res3b & FP16, b=8 & 5.80 & 616.86 (0.46) & 8.45 & 420.72 (0.37) & 6.75 & 533.04 (0.61)\\
    TX2 &  res3b & FP16, b=16 & 11.26 & 629.89 (0.47) & 16.50 & 426.74 (0.37) & 12.95 & 544.73 (0.62)\\
    TX2 &  res3b & FP16, b=32 & 22.24 & 630.78 (0.47) & 33.51 & 428.78 (0.37) & 25.61 & 547.40 (0.63)\\
    TX2 &  res3b & FP16, b=64 & 44.38 & 628.12 (0.47) & 65.79 & 434.20 (0.38) & 51.07 & 546.72 (0.63)\\
    TX2 &  res3b & FP16, b=128 & 87.72 & 636.16 (0.48) & 129.87 & 434.62 (0.38) & 101.77 & 548.74 (0.63)\\ \midrule
    TX2 &  res3c & FP16, b=1 & 1.05 & 506.48 (0.38) & 1.42 & 343.22 (0.30) & 1.22 & 436.74 (0.50)\\
    TX2 &  res3c & FP16, b=2 & 1.70 & 561.84 (0.42) & 2.43 & 380.61 (0.33) & 1.96 & 486.64 (0.56)\\
    TX2 &  res3c & FP16, b=4 & 3.06 & 598.69 (0.45) & 4.46 & 406.57 (0.35) & 3.56 & 515.82 (0.59)\\
    TX2 &  res3c & FP16, b=8 & 5.81 & 613.47 (0.46) & 8.47 & 419.93 (0.37) & 6.70 & 536.24 (0.61)\\
    TX2 &  res3c & FP16, b=16 & 11.19 & 630.78 (0.47) & 16.40 & 428.78 (0.37) & 12.97 & 541.43 (0.62)\\
    TX2 &  res3c & FP16, b=32 & 22.05 & 632.56 (0.47) & 32.55 & 430.43 (0.37) & 25.68 & 546.72 (0.63)\\
    TX2 &  res3c & FP16, b=64 & 43.94 & 637.07 (0.48) & 64.34 & 434.20 (0.38) & 50.46 & 552.13 (0.63)\\
    TX2 &  res3c & FP16, b=128 & 87.87 & 636.16 (0.48) & 129.02 & 431.68 (0.38) & 101.34 & 550.09 (0.63)\\ \midrule
    TX2 &  res3d & FP16, b=1 & 1.04 & 504.19 (0.38) & 1.40 & 347.76 (0.30) & 1.22 & 438.45 (0.50)\\
    TX2 &  res3d & FP16, b=2 & 1.70 & 561.13 (0.42) & 2.40 & 385.20 (0.34) & 1.95 & 488.77 (0.56)\\
    TX2 &  res3d & FP16, b=4 & 3.08 & 592.35 (0.44) & 4.43 & 408.05 (0.36) & 3.55 & 515.23 (0.59)\\
    TX2 &  res3d & FP16, b=8 & 5.84 & 613.47 (0.46) & 8.41 & 423.10 (0.37) & 6.71 & 536.88 (0.61)\\
    TX2 &  res3d & FP16, b=16 & 11.25 & 624.61 (0.47) & 16.48 & 427.15 (0.37) & 12.93 & 546.72 (0.63)\\
    TX2 &  res3d & FP16, b=32 & 22.04 & 631.67 (0.47) & 32.74 & 430.85 (0.37) & 25.51 & 551.45 (0.63)\\
    TX2 &  res3d & FP16, b=64 & 44.27 & 633.46 (0.48) & 64.21 & 435.89 (0.38) & 50.51 & 553.49 (0.63)\\
    TX2 &  res3d & FP16, b=128 & 88.66 & 630.78 (0.47) & 129.91 & 431.27 (0.38) & 101.91 & 549.41 (0.63)\\ \midrule
    TX2 &  res4a & FP16, b=1 & 1.20 & 575.15 (0.43) & 1.66 & 386.37 (0.34) & 1.39 & 496.77 (0.57)\\
    TX2 &  res4a & FP16, b=2 & 2.09 & 604.46 (0.45) & 3.01 & 411.33 (0.36) & 2.41 & 524.32 (0.60)\\
    TX2 &  res4a & FP16, b=4 & 3.74 & 654.12 (0.49) & 5.46 & 446.87 (0.39) & 4.34 & 569.20 (0.65)\\
    TX2 &  res4a & FP16, b=8 & 7.00 & 688.35 (0.52) & 10.16 & 472.33 (0.41) & 8.03 & 600.26 (0.69)\\
    TX2 &  res4a & FP16, b=16 & 13.40 & 706.02 (0.53) & 19.71 & 480.59 (0.42) & 15.37 & 621.85 (0.71)\\
    TX2 &  res4a & FP16, b=32 & 26.40 & 713.52 (0.54) & 38.60 & 493.12 (0.43) & 30.30 & 627.01 (0.72)\\
    TX2 &  res4a & FP16, b=64 & 52.32 & 722.89 (0.54) & 76.46 & 494.33 (0.43) & 60.23 & 628.96 (0.72)\\
    TX2 &  res4a & FP16, b=128 & 104.66 & 723.76 (0.54) & 152.47 & 496.77 (0.43) & 119.12 & 633.57 (0.72)\\
			\midrule
			\midrule
		\end{tabular}
	}
\vspace*{-3ex}
\end{table}

\begin{table}[!ht]
	\resizebox{0.7\width}{!}{
		\begin{tabular}{|lll||cc|cc|cc|}
			\toprule
			 & & &  \multicolumn{2}{c|}{\textbf{MaxN}} & \multicolumn{2}{c|}{\textbf{MaxQ}}  & \multicolumn{2}{c|}{\textbf{MaxP}}\\
			\textbf{HW} & \textbf{Layer} & \textbf{Parameters} & \textbf{Lat} &  \textbf{Throughput (Eff)} & \textbf{Lat} & \textbf{Throughput (Eff)} & \textbf{Lat} &  \textbf{Throughput (Eff)}\\
			 & & & \textbf{[ms]} & \textbf{[GOPs] ([\%])} & \textbf{[ms]} & \textbf{[GOPs] ([\%])} & \textbf{[ms]} & \textbf{[GOPs] ([\%])}  \\
			\midrule
    TX2 &  res4b & FP16, b=1 & 1.03 & 599.49 (0.45) & 1.40 & 407.31 (0.35) & 1.16 & 514.64 (0.59)\\
    TX2 &  res4b & FP16, b=2 & 1.53 & 644.41 (0.48) & 2.11 & 437.60 (0.38) & 1.74 & 562.54 (0.64)\\
    TX2 &  res4b & FP16, b=4 & 2.60 & 706.51 (0.53) & 3.81 & 474.76 (0.41) & 3.03 & 610.96 (0.70)\\
    TX2 &  res4b & FP16, b=8 & 4.87 & 740.43 (0.56) & 7.16 & 497.46 (0.43) & 5.58 & 648.15 (0.74)\\
    TX2 &  res4b & FP16, b=16 & 9.33 & 763.18 (0.57) & 13.65 & 520.03 (0.45) & 10.82 & 660.60 (0.76)\\
    TX2 &  res4b & FP16, b=32 & 18.13 & 769.75 (0.58) & 26.64 & 526.15 (0.46) & 20.84 & 674.54 (0.77)\\
    TX2 &  res4b & FP16, b=64 & 36.31 & 776.43 (0.58) & 52.95 & 530.51 (0.46) & 41.50 & 675.56 (0.77)\\
    TX2 &  res4b & FP16, b=128 & 71.33 & 780.49 (0.59) & 105.55 & 529.88 (0.46) & 82.15 & 680.70 (0.78)\\ \midrule
    TX2 &  res4c & FP16, b=1 & 1.03 & 598.69 (0.45) & 1.40 & 408.42 (0.36) & 1.16 & 511.11 (0.58)\\
    TX2 &  res4c & FP16, b=2 & 1.52 & 647.21 (0.49) & 2.12 & 437.60 (0.38) & 1.74 & 557.63 (0.64)\\
    TX2 &  res4c & FP16, b=4 & 2.63 & 695.53 (0.52) & 3.79 & 474.76 (0.41) & 3.06 & 602.72 (0.69)\\
    TX2 &  res4c & FP16, b=8 & 5.17 & 697.69 (0.52) & 7.11 & 499.69 (0.43) & 5.61 & 644.41 (0.74)\\
    TX2 &  res4c & FP16, b=16 & 9.29 & 764.48 (0.57) & 13.64 & 518.82 (0.45) & 10.74 & 660.60 (0.76)\\
    TX2 &  res4c & FP16, b=32 & 18.15 & 773.74 (0.58) & 26.63 & 528.63 (0.46) & 21.05 & 666.50 (0.76)\\
    TX2 &  res4c & FP16, b=64 & 35.81 & 780.49 (0.59) & 52.86 & 530.51 (0.46) & 41.59 & 674.54 (0.77)\\
    TX2 &  res4c & FP16, b=128 & 72.52 & 773.74 (0.58) & 105.65 & 527.39 (0.46) & 82.21 & 680.70 (0.78)\\ \midrule
    TX2 &  res4d & FP16, b=1 & 1.02 & 599.49 (0.45) & 1.40 & 408.05 (0.36) & 1.16 & 514.64 (0.59)\\
    TX2 &  res4d & FP16, b=2 & 1.52 & 650.98 (0.49) & 2.12 & 438.02 (0.38) & 1.74 & 558.33 (0.64)\\
    TX2 &  res4d & FP16, b=4 & 2.62 & 703.18 (0.53) & 3.82 & 474.26 (0.41) & 3.02 & 613.47 (0.70)\\
    TX2 &  res4d & FP16, b=8 & 4.84 & 745.37 (0.56) & 7.13 & 503.06 (0.44) & 5.60 & 644.41 (0.74)\\
    TX2 &  res4d & FP16, b=16 & 9.29 & 763.18 (0.57) & 13.58 & 521.24 (0.45) & 10.72 & 661.57 (0.76)\\
    TX2 &  res4d & FP16, b=32 & 18.32 & 769.75 (0.58) & 26.49 & 529.88 (0.46) & 20.92 & 673.53 (0.77)\\
    TX2 &  res4d & FP16, b=64 & 36.17 & 775.08 (0.58) & 53.11 & 528.63 (0.46) & 41.41 & 678.64 (0.78)\\
    TX2 &  res4d & FP16, b=128 & 72.01 & 779.13 (0.58) & 105.92 & 526.15 (0.46) & 82.10 & 679.67 (0.78)\\ \midrule
    TX2 &  res4e & FP16, b=1 & 1.02 & 601.10 (0.45) & 1.40 & 407.31 (0.35) & 1.16 & 514.64 (0.59)\\
    TX2 &  res4e & FP16, b=2 & 1.53 & 649.09 (0.49) & 2.12 & 437.17 (0.38) & 1.75 & 560.43 (0.64)\\
    TX2 &  res4e & FP16, b=4 & 2.64 & 696.61 (0.52) & 3.81 & 474.26 (0.41) & 3.03 & 610.13 (0.70)\\
    TX2 &  res4e & FP16, b=8 & 4.85 & 740.43 (0.56) & 7.19 & 500.81 (0.44) & 5.63 & 644.41 (0.74)\\
    TX2 &  res4e & FP16, b=16 & 9.33 & 758.00 (0.57) & 13.67 & 518.22 (0.45) & 10.68 & 663.53 (0.76)\\
    TX2 &  res4e & FP16, b=32 & 18.10 & 771.07 (0.58) & 26.59 & 526.15 (0.46) & 20.95 & 671.51 (0.77)\\
    TX2 &  res4e & FP16, b=64 & 35.83 & 780.49 (0.59) & 52.81 & 531.14 (0.46) & 41.24 & 676.58 (0.77)\\
    TX2 &  res4e & FP16, b=128 & 72.39 & 775.08 (0.58) & 105.40 & 529.25 (0.46) & 82.50 & 679.67 (0.78)\\ \midrule
    TX2 &  res4f & FP16, b=1 & 1.02 & 599.49 (0.45) & 1.39 & 408.42 (0.36) & 1.17 & 514.05 (0.59)\\
    TX2 &  res4f & FP16, b=2 & 1.51 & 650.03 (0.49) & 2.12 & 437.17 (0.38) & 1.75 & 559.03 (0.64)\\
    TX2 &  res4f & FP16, b=4 & 2.64 & 699.88 (0.53) & 3.80 & 474.76 (0.41) & 3.02 & 610.96 (0.70)\\
    TX2 &  res4f & FP16, b=8 & 4.83 & 741.66 (0.56) & 7.10 & 500.81 (0.44) & 5.60 & 642.56 (0.74)\\
    TX2 &  res4f & FP16, b=16 & 9.27 & 763.18 (0.57) & 13.57 & 521.85 (0.45) & 10.76 & 664.52 (0.76)\\
    TX2 &  res4f & FP16, b=32 & 18.24 & 767.10 (0.58) & 26.66 & 525.52 (0.46) & 21.02 & 669.49 (0.77)\\
    TX2 &  res4f & FP16, b=64 & 36.10 & 771.07 (0.58) & 52.77 & 532.41 (0.46) & 41.20 & 675.56 (0.77)\\
    TX2 &  res4f & FP16, b=128 & 71.76 & 776.43 (0.58) & 105.63 & 533.68 (0.46) & 82.10 & 677.61 (0.78)\\ \midrule
    TX2 &  res5a & FP16, b=1 & 1.73 & 413.29 (0.31) & 2.43 & 281.55 (0.25) & 1.98 & 357.60 (0.41)\\
    TX2 &  res5a & FP16, b=2 & 2.05 & 634.24 (0.48) & 2.88 & 430.04 (0.37) & 2.34 & 552.57 (0.63)\\
    TX2 &  res5a & FP16, b=4 & 3.68 & 664.17 (0.50) & 5.44 & 447.20 (0.39) & 4.29 & 576.80 (0.66)\\
    TX2 &  res5a & FP16, b=8 & 7.30 & 659.82 (0.49) & 10.75 & 445.88 (0.39) & 8.45 & 572.43 (0.65)\\
    TX2 &  res5a & FP16, b=16 & 13.52 & 707.67 (0.53) & 19.64 & 484.05 (0.42) & 15.51 & 615.53 (0.70)\\
    TX2 &  res5a & FP16, b=32 & 25.37 & 746.07 (0.56) & 36.80 & 514.95 (0.45) & 29.04 & 656.96 (0.75)\\
    TX2 &  res5a & FP16, b=64 & 48.45 & 778.71 (0.58) & 71.37 & 532.16 (0.46) & 56.06 & 675.29 (0.77)\\
    TX2 &  res5a & FP16, b=128 & 95.61 & 791.96 (0.59) & 140.12 & 540.72 (0.47) & 110.13 & 686.01 (0.78)\\ \midrule
    TX2 &  res5b & FP16, b=1 & 1.24 & 456.82 (0.34) & 1.71 & 307.16 (0.27) & 1.41 & 391.96 (0.45)\\
    TX2 &  res5b & FP16, b=2 & 1.60 & 666.50 (0.50) & 2.29 & 447.22 (0.39) & 1.84 & 572.63 (0.66)\\
    TX2 &  res5b & FP16, b=4 & 2.66 & 704.29 (0.53) & 3.85 & 472.75 (0.41) & 3.12 & 607.64 (0.70)\\
    TX2 &  res5b & FP16, b=8 & 4.98 & 723.66 (0.54) & 7.40 & 481.40 (0.42) & 5.79 & 626.36 (0.72)\\
    TX2 &  res5b & FP16, b=16 & 8.79 & 810.18 (0.61) & 13.01 & 540.78 (0.47) & 10.17 & 704.29 (0.81)\\
    TX2 &  res5b & FP16, b=32 & 16.33 & 861.70 (0.65) & 24.27 & 577.06 (0.50) & 18.76 & 752.90 (0.86)\\
    TX2 &  res5b & FP16, b=64 & 31.47 & 892.66 (0.67) & 47.00 & 596.29 (0.52) & 36.27 & 775.08 (0.89)\\
    TX2 &  res5b & FP16, b=128 & 61.55 & 907.14 (0.68) & 92.43 & 606.82 (0.53) & 71.35 & 787.36 (0.90)\\ \midrule
    TX2 &  res5c & FP16, b=1 & 1.23 & 460.58 (0.35) & 1.71 & 312.74 (0.27) & 1.38 & 398.95 (0.46)\\
    TX2 &  res5c & FP16, b=2 & 1.59 & 665.51 (0.50) & 2.28 & 448.57 (0.39) & 1.84 & 571.89 (0.65)\\
    TX2 &  res5c & FP16, b=4 & 2.66 & 705.40 (0.53) & 3.85 & 469.77 (0.41) & 3.10 & 610.13 (0.70)\\
    TX2 &  res5c & FP16, b=8 & 4.99 & 720.16 (0.54) & 7.48 & 480.37 (0.42) & 5.80 & 626.36 (0.72)\\
    TX2 &  res5c & FP16, b=16 & 8.79 & 811.66 (0.61) & 13.08 & 540.78 (0.47) & 10.11 & 703.18 (0.80)\\
    TX2 &  res5c & FP16, b=32 & 16.42 & 863.36 (0.65) & 24.29 & 576.32 (0.50) & 18.80 & 746.62 (0.85)\\
    TX2 &  res5c & FP16, b=64 & 31.34 & 892.66 (0.67) & 47.02 & 596.29 (0.52) & 36.12 & 777.78 (0.89)\\
    TX2 &  res5c & FP16, b=128 & 61.58 & 907.14 (0.68) & 92.32 & 603.54 (0.53) & 71.16 & 787.36 (0.90)\\
			\midrule
			\midrule
		\end{tabular}
	}
\vspace*{-3ex}
\end{table}

\begin{table}[!ht]
	\caption{Level 3 - Inference Results ResNet50}
	\label{AT_L3R}
	\resizebox{0.7\width}{!}{
		\begin{tabular}{|ll|c|cc|cc|c|}
			\toprule
			\textbf{ResNet50} &  & \textbf{Top-5 (Top-1) Acc} & 
			\multicolumn{2}{c|}{\textbf{Latency [ms]}} & \multicolumn{2}{c|}{\textbf{Throughput [GOPs] (Efficiency [\%])}} & \textbf{Power [W]}  \\
			 \textbf{Platform} & \textbf{Parameters} & \textbf{[\%]} & \textbf{system} & \textbf{compute} & \textbf{system} & \textbf{compute}  & \\
			\midrule
ZCU104 & INT8, t=1 & 90.85 (72.53) & 17.96 & 14.96 & 324.91 (0.08) & 516.07 (0.13) & 21.41\\
ZCU104 & INT8, t=2 & 90.85 (72.53) & 19.69 & 16.63 & 786.49 (0.19) & 931.12 (0.23) & 25.78\\
ZCU104 & INT8, t=3 & 90.85 (72.53) & 25.37 & 22.46 & 906.78 (0.22) & 1029.32 (0.25) & 26.39\\
ZCU104 & INT8, t=4 & 90.85 (72.53) & 33.46 & 30.48 & 918.25 (0.23) & 1017.81 (0.25) & 26.53\\
ZCU104 & INT8, t=5 & 90.85 (72.53) & 40.28 & 37.26 & 943.85 (0.23) & 1067.31 (0.26) & 26.82\\
ZCU104 & INT8, t=6 & 90.85 (72.53) & 47.95 & 45.27 & 946.97 (0.23) & 1056.38 (0.26) & 26.83\\
ZCU104 & INT8, t=7 & 90.85 (72.53) & 56.74 & 53.61 & 943.37 (0.23) & 1047.84 (0.26) & 26.86\\
ZCU104 & INT8, t=8 & 90.85 (72.53) & 64.88 & 62.30 & 948.05 (0.23) & 1044.11 (0.26) & 26.89\\
\midrule
\midrule
TX2, MaxN & FP16, b=1 & 92.12 (75.11) & 13.99 & 10.68 & 547.84 (0.41) & 725.23 (0.54) & 12.57\\
TX2, MaxN & FP16, b=2 & 92.12 (75.11) & 23.65 & 18.25 & 651.39 (0.49) & 855.59 (0.64) & 13.57\\
TX2, MaxN & FP16, b=4 & 92.12 (75.11) & 43.79 & 34.23 & 703.76 (0.53) & 911.67 (0.68) & 13.76\\
TX2, MaxN & FP16, b=8 & 92.12 (75.11) & 84.79 & 65.86 & 728.19 (0.55) & 937.28 (0.70) & 13.92\\
TX2, MaxN & FP16, b=16 & 92.12 (75.11) & 162.58 & 126.23 & 759.76 (0.57) & 976.60 (0.73) & 13.69\\
TX2, MaxN & FP16, b=32 & 92.12 (75.11) & 317.87 & 247.34 & 779.00 (0.58) & 999.84 (0.75) & 13.70\\
TX2, MaxN & FP16, b=64 & 92.12 (75.11) & 620.08 & 490.37 & 799.24 (0.60) & 1006.85 (0.76) & 13.76\\
TX2, MaxN & FP16, b=128 & 92.12 (75.11) & 1211.85 & 975.98 & 809.47 (0.61) & 1011.95 (0.76) & 13.82\\
TX2, MaxN & FP32, b=1 & 92.11 (75.15) & 22.32 & 18.97 & 344.88 (0.26) & 407.51 (0.31) & 14.58\\
TX2, MaxN & FP32, b=2 & 92.11 (75.15) & 38.46 & 32.96 & 401.14 (0.30) & 470.87 (0.35) & 15.02\\
TX2, MaxN & FP32, b=4 & 92.11 (75.15) & 72.96 & 62.96 & 423.04 (0.32) & 491.63 (0.37) & 15.05\\
TX2, MaxN & FP32, b=8 & 92.11 (75.15) & 141.13 & 122.18 & 437.53 (0.33) & 506.07 (0.38) & 15.19\\
TX2, MaxN & FP32, b=16 & 92.11 (75.15) & 272.41 & 235.85 & 453.44 (0.34) & 523.42 (0.39) & 15.34\\
TX2, MaxN & FP32, b=32 & 92.11 (75.15) & 531.12 & 460.67 & 465.45 (0.35) & 536.20 (0.40) & 15.39\\
TX2, MaxN & FP32, b=64 & 92.11 (75.15) & 1042.67 & 913.42 & 473.23 (0.36) & 539.88 (0.41) & 15.51\\
TX2, MaxN & FP32, b=128 & 92.11 (75.15) & 2115.54 & 1810.90 & 462.78 (0.35) & 544.51 (0.41) & 15.21\\
\midrule
TX2, MaxQ & FP16, b=1 & 92.12 (75.11) & 20.46 & 15.86 & 376.05 (0.28) & 496.32 (0.37) & 6.83\\
TX2, MaxQ & FP16, b=2 & 92.12 (75.11) & 34.65 & 26.69 & 444.47 (0.33) & 582.64 (0.44) & 6.94\\
TX2, MaxQ & FP16, b=4 & 92.12 (75.11) & 64.53 & 50.01 & 477.72 (0.36) & 618.73 (0.46) & 7.00\\
TX2, MaxQ & FP16, b=8 & 92.12 (75.11) & 124.75 & 96.69 & 494.70 (0.37) & 638.71 (0.48) & 7.06\\
TX2, MaxQ & FP16, b=16 & 92.12 (75.11) & 239.00 & 185.17 & 516.18 (0.39) & 666.23 (0.50) & 7.13\\
TX2, MaxQ & FP16, b=32 & 92.12 (75.11) & 466.49 & 362.02 & 529.31 (0.40) & 682.18 (0.51) & 7.12\\
TX2, MaxQ & FP16, b=64 & 92.12 (75.11) & 924.53 & 717.12 & 534.18 (0.40) & 687.93 (0.52) & 7.13\\
TX2, MaxQ & FP16, b=128 & 92.12 (75.11) & 1838.48 & 1429.00 & 536.31 (0.40) & 691.03 (0.52) & 7.15\\
TX2, MaxQ & FP32, b=1 & 92.11 (75.15) & 32.66 & 27.94 & 235.37 (0.18) & 279.20 (0.21) & 7.60\\
TX2, MaxQ & FP32, b=2 & 92.11 (75.15) & 56.36 & 48.32 & 273.37 (0.21) & 320.87 (0.24) & 7.79\\
TX2, MaxQ & FP32, b=4 & 92.11 (75.15) & 106.84 & 92.09 & 288.78 (0.22) & 335.67 (0.25) & 7.77\\
TX2, MaxQ & FP32, b=8 & 92.11 (75.15) & 207.45 & 179.45 & 297.69 (0.22) & 344.35 (0.26) & 7.85\\
TX2, MaxQ & FP32, b=16 & 92.11 (75.15) & 398.74 & 344.35 & 309.54 (0.23) & 358.82 (0.27) & 7.97\\
TX2, MaxQ & FP32, b=32 & 92.11 (75.15) & 779.69 & 673.93 & 316.63 (0.24) & 366.42 (0.27) & 7.99\\
TX2, MaxQ & FP32, b=64 & 92.11 (75.15) & 1540.33 & 1333.24 & 320.57 (0.24) & 370.36 (0.28) & 8.03\\
TX2, MaxQ & FP32, b=128 & 92.11 (75.15) & 3118.09 & 2650.98 & 315.38 (0.24) & 372.08 (0.28) & 7.93\\
\midrule
TX2, MaxP & FP16, b=1 & 92.12 (75.11) & 16.52 & 12.46 & 464.14 (0.35) & 632.05 (0.47) & 9.38\\
TX2, MaxP & FP16, b=2 & 92.12 (75.11) & 28.10 & 20.92 & 547.57 (0.41) & 745.09 (0.56) & 9.59\\
TX2, MaxP & FP16, b=4 & 92.12 (75.11) & 52.22 & 39.14 & 590.78 (0.44) & 790.58 (0.59) & 9.71\\
TX2, MaxP & FP16, b=8 & 92.12 (75.11) & 100.21 & 75.54 & 615.58 (0.46) & 818.10 (0.61) & 9.81\\
TX2, MaxP & FP16, b=16 & 92.12 (75.11) & 193.89 & 145.36 & 637.01 (0.48) & 849.79 (0.64) & 9.79\\
TX2, MaxP & FP16, b=32 & 92.12 (75.11) & 375.86 & 283.47 & 657.79 (0.49) & 870.94 (0.65) & 9.79\\
TX2, MaxP & FP16, b=64 & 92.12 (75.11) & 733.74 & 562.74 & 673.19 (0.51) & 879.15 (0.66) & 9.81\\
TX2, MaxP & FP16, b=128 & 92.12 (75.11) & 1453.87 & 1121.27 & 677.63 (0.51) & 879.65 (0.66) & 9.86\\
TX2, MaxP & FP32, b=1 & 92.11 (75.15) & 26.16 & 22.00 & 293.60 (0.22) & 355.65 (0.27) & 10.54\\
TX2, MaxP & FP32, b=2 & 92.11 (75.15) & 45.07 & 37.86 & 341.72 (0.26) & 410.11 (0.31) & 10.83\\
TX2, MaxP & FP32, b=4 & 92.11 (75.15) & 85.31 & 72.25 & 361.93 (0.27) & 428.79 (0.32) & 10.86\\
TX2, MaxP & FP32, b=8 & 92.11 (75.15) & 165.16 & 140.36 & 373.86 (0.28) & 440.83 (0.33) & 11.01\\
TX2, MaxP & FP32, b=16 & 92.11 (75.15) & 318.38 & 270.54 & 387.72 (0.29) & 456.71 (0.34) & 11.13\\
TX2, MaxP & FP32, b=32 & 92.11 (75.15) & 621.30 & 528.54 & 397.67 (0.30) & 467.21 (0.35) & 11.15\\
TX2, MaxP & FP32, b=64 & 92.11 (75.15) & 1219.10 & 1046.82 & 404.38 (0.30) & 471.56 (0.35) & 11.24\\
TX2, MaxP & FP32, b=128 & 92.11 (75.15) & 2495.53 & 2076.54 & 393.22 (0.29) & 474.95 (0.36) & 10.88\\
            \midrule
            \midrule
		\end{tabular}
	}
	\vspace*{-3ex}
\end{table}

\begin{table}[!ht]
	\caption{Level 3 - Inference Results GoogleNetV1}
	\label{AT_L3G}
	\resizebox{0.7\width}{!}{
		\begin{tabular}{|ll|c|cc|cc|c|}
			\toprule
			\textbf{GoogLeNet} &  & \textbf{Top-5 (Top-1) Acc} & 
			\multicolumn{2}{c|}{\textbf{Latency [ms]}} & \multicolumn{2}{c|}{\textbf{Throughput [GOPs] (Efficiency [\%])}} & \textbf{Power [W]}  \\
			 \textbf{Platform} & \textbf{Parameters} & \textbf{[\%]} & \textbf{system} & \textbf{compute} & \textbf{system} & \textbf{compute}  & \\
			\midrule
ZCU104 & INT8, t=1 & 89.26 (69.49) & 9.65 & 6.68 & 323.50 (0.08) & 468.64 (0.12) & 21.49\\
ZCU104 & INT8, t=2 & 89.26 (69.49) & 9.99 & 7.06 & 499.97 (0.12) & 895.19 (0.22) & 24.60\\
ZCU104 & INT8, t=3 & 89.26 (69.49) & 11.88 & 9.30 & 784.16 (0.19) & 1050.36 (0.26) & 25.41\\
ZCU104 & INT8, t=4 & 89.26 (69.49) & 15.85 & 12.95 & 782.99 (0.19) & 971.12 (0.24) & 25.50\\
ZCU104 & INT8, t=5 & 89.26 (69.49) & 18.24 & 15.27 & 848.31 (0.21) & 1122.66 (0.28) & 25.96\\
ZCU104 & INT8, t=6 & 89.26 (69.49) & 21.88 & 18.35 & 851.75 (0.21) & 1116.67 (0.27) & 26.00\\
ZCU104 & INT8, t=7 & 89.26 (69.49) & 25.60 & 22.78 & 853.81 (0.21) & 1081.32 (0.27) & 25.93\\
ZCU104 & INT8, t=8 & 89.26 (69.49) & 28.91 & 25.94 & 846.70 (0.21) & 1062.84 (0.26) & 25.90\\
\midrule
\midrule
TX2, MaxN & FP16, b=1 & 87.85 (66.94) & 7.98 & 5.17 & 388.92 (0.29) & 612.39 (0.46) & 11.52\\
TX2, MaxN & FP16, b=2 & 87.85 (66.94) & 14.68 & 9.23 & 438.58 (0.33) & 693.35 (0.52) & 11.40\\
TX2, MaxN & FP16, b=4 & 87.85 (66.94) & 26.31 & 16.36 & 475.75 (0.36) & 770.88 (0.58) & 12.00\\
TX2, MaxN & FP16, b=8 & 87.85 (66.94) & 49.83 & 30.96 & 508.99 (0.38) & 811.73 (0.61) & 12.21\\
TX2, MaxN & FP16, b=16 & 87.85 (66.94) & 95.81 & 59.47 & 544.72 (0.41) & 843.35 (0.63) & 12.34\\
TX2, MaxN & FP16, b=32 & 87.85 (66.94) & 185.96 & 116.62 & 538.49 (0.40) & 859.48 (0.64) & 11.85\\
TX2, MaxN & FP16, b=64 & 87.85 (66.94) & 389.12 & 231.45 & 511.92 (0.38) & 862.22 (0.65) & 11.62\\
TX2, MaxN & FP16, b=128 & 87.85 (66.94) & 810.86 & 459.85 & 493.02 (0.37) & 871.35 (0.65) & 11.29\\
TX2, MaxN & FP32, b=1 & 87.84 (66.94) & 12.01 & 8.71 & 259.52 (0.19) & 361.81 (0.27) & 12.79\\
TX2, MaxN & FP32, b=2 & 87.84 (66.94) & 21.01 & 15.62 & 297.38 (0.22) & 406.52 (0.30) & 13.06\\
TX2, MaxN & FP32, b=4 & 87.84 (66.94) & 38.37 & 28.47 & 325.96 (0.24) & 441.59 (0.33) & 13.35\\
TX2, MaxN & FP32, b=8 & 87.84 (66.94) & 72.78 & 53.96 & 320.83 (0.24) & 464.59 (0.35) & 13.56\\
TX2, MaxN & FP32, b=16 & 87.84 (66.94) & 141.92 & 104.99 & 353.06 (0.26) & 477.28 (0.36) & 13.70\\
TX2, MaxN & FP32, b=32 & 87.84 (66.94) & 279.20 & 206.82 & 361.42 (0.27) & 484.49 (0.36) & 13.77\\
TX2, MaxN & FP32, b=64 & 87.84 (66.94) & 571.52 & 409.48 & 347.85 (0.26) & 488.15 (0.37) & 13.53\\
TX2, MaxN & FP32, b=128 & 87.84 (66.94) & 1165.91 & 811.57 & 342.21 (0.26) & 490.41 (0.37) & 13.17\\
\midrule
TX2, MaxQ & FP16, b=1 & 87.85 (66.94) & 12.42 & 7.71 & 249.66 (0.19) & 420.01 (0.32) & 5.88\\
TX2, MaxQ & FP16, b=2 & 87.85 (66.94) & 21.48 & 13.39 & 290.21 (0.22) & 474.86 (0.36) & 6.01\\
TX2, MaxQ & FP16, b=4 & 87.85 (66.94) & 38.55 & 23.99 & 323.91 (0.24) & 524.72 (0.39) & 6.13\\
TX2, MaxQ & FP16, b=8 & 87.85 (66.94) & 73.30 & 45.40 & 340.72 (0.26) & 552.66 (0.41) & 6.20\\
TX2, MaxQ & FP16, b=16 & 87.85 (66.94) & 141.55 & 87.55 & 353.38 (0.27) & 571.79 (0.43) & 6.24\\
TX2, MaxQ & FP16, b=32 & 87.85 (66.94) & 277.12 & 172.07 & 362.00 (0.27) & 582.64 (0.44) & 6.24\\
TX2, MaxQ & FP16, b=64 & 87.85 (66.94) & 574.23 & 340.85 & 346.43 (0.26) & 588.26 (0.44) & 6.09\\
TX2, MaxQ & FP16, b=128 & 87.85 (66.94) & 1182.94 & 680.24 & 335.61 (0.25) & 589.83 (0.44) & 6.03\\
TX2, MaxQ & FP32, b=1 & 87.84 (66.94) & 17.54 & 12.82 & 177.36 (0.13) & 249.47 (0.19) & 6.70\\
TX2, MaxQ & FP32, b=2 & 87.84 (66.94) & 30.80 & 22.77 & 202.46 (0.15) & 277.85 (0.21) & 6.89\\
TX2, MaxQ & FP32, b=4 & 87.84 (66.94) & 56.49 & 41.81 & 220.94 (0.17) & 300.29 (0.23) & 7.02\\
TX2, MaxQ & FP32, b=8 & 87.84 (66.94) & 107.28 & 79.34 & 233.39 (0.18) & 316.27 (0.24) & 7.11\\
TX2, MaxQ & FP32, b=16 & 87.84 (66.94) & 208.77 & 154.43 & 239.76 (0.18) & 324.46 (0.24) & 7.16\\
TX2, MaxQ & FP32, b=32 & 87.84 (66.94) & 410.81 & 303.79 & 244.50 (0.18) & 330.06 (0.25) & 7.18\\
TX2, MaxQ & FP32, b=64 & 87.84 (66.94) & 835.95 & 602.25 & 238.58 (0.18) & 332.87 (0.25) & 7.08\\
TX2, MaxQ & FP32, b=128 & 87.84 (66.94) & 1702.86 & 1196.52 & 232.88 (0.17) & 333.96 (0.25) & 7.02\\
\midrule
TX2, MaxP & FP16, b=1 & 87.85 (66.94) & 9.69 & 6.09 & 318.13 (0.24) & 532.93 (0.40) & 8.07\\
TX2, MaxP & FP16, b=2 & 87.85 (66.94) & 16.66 & 10.53 & 374.39 (0.28) & 606.07 (0.45) & 8.28\\
TX2, MaxP & FP16, b=4 & 87.85 (66.94) & 31.93 & 18.85 & 415.17 (0.31) & 668.51 (0.50) & 8.47\\
TX2, MaxP & FP16, b=8 & 87.85 (66.94) & 60.46 & 35.59 & 428.38 (0.32) & 705.81 (0.53) & 8.61\\
TX2, MaxP & FP16, b=16 & 87.85 (66.94) & 108.66 & 68.56 & 459.58 (0.34) & 731.15 (0.55) & 8.69\\
TX2, MaxP & FP16, b=32 & 87.85 (66.94) & 226.68 & 134.27 & 443.19 (0.33) & 745.48 (0.56) & 8.35\\
TX2, MaxP & FP16, b=64 & 87.85 (66.94) & 478.58 & 266.78 & 409.86 (0.31) & 749.71 (0.56) & 8.07\\
TX2, MaxP & FP16, b=128 & 87.85 (66.94) & 1019.48 & 530.68 & 392.21 (0.29) & 753.68 (0.57) & 7.74\\
TX2, MaxP & FP32, b=1 & 87.84 (66.94) & 14.35 & 10.17 & 217.30 (0.16) & 315.68 (0.24) & 9.15\\
TX2, MaxP & FP32, b=2 & 87.84 (66.94) & 25.01 & 17.89 & 249.21 (0.19) & 354.04 (0.27) & 9.37\\
TX2, MaxP & FP32, b=4 & 87.84 (66.94) & 45.84 & 32.72 & 272.61 (0.20) & 382.77 (0.29) & 9.56\\
TX2, MaxP & FP32, b=8 & 87.84 (66.94) & 86.85 & 61.95 & 290.61 (0.22) & 404.57 (0.30) & 9.71\\
TX2, MaxP & FP32, b=16 & 87.84 (66.94) & 169.26 & 120.61 & 295.84 (0.22) & 415.80 (0.31) & 9.75\\
TX2, MaxP & FP32, b=32 & 87.84 (66.94) & 330.80 & 237.45 & 302.80 (0.23) & 421.99 (0.32) & 9.77\\
TX2, MaxP & FP32, b=64 & 87.84 (66.94) & 687.11 & 470.42 & 288.46 (0.22) & 425.71 (0.32) & 9.50\\
TX2, MaxP & FP32, b=128 & 87.84 (66.94) & 1426.28 & 932.76 & 279.73 (0.21) & 427.71 (0.32) & 9.16\\
            \midrule
            \midrule
		\end{tabular}
	}
	\vspace*{-3ex}
\end{table}

\small{
\begin{table}[!ht]
\caption{Level 1 and 2 - Discrepancy between latency of different convolutions and residual layers}
\label{l1-var}
\resizebox{\linewidth}{!}{
\begin{tabular}{|l|l|cc||l|l|cc|cc|}
\toprule
\multicolumn{4}{|c||}{\textbf{Level 2}} & \multicolumn{6}{c|}{\textbf{Level 1}} \\
\midrule
\textbf{Residual} &  & \multicolumn{2}{c||}{\textbf{TX2, MaxN, FP16}}  & \textbf{Conv.} &   & \multicolumn{2}{c|}{\textbf{TX2, MaxN, FP16}} & \multicolumn{2}{c|}{\textbf{ZCU104,INT8}}  \\
\textbf{Layer} & \textbf{[MOP]} & \textbf{b=1 [ms]} & \textbf{b=128 [ms]} & \textbf{Layer} & \textbf{[MOP]} & \textbf{b=1 [ms]} & \textbf{b=128 [ms]} & \textbf{t=1 [ms]} & \textbf{t=8 [ms]} \\
\midrule
res2a & 462.44 & 1.37 & 119.12 & res2a\_branch2a,  1x1 & 25.70 & 0.06 & 5.05 & 0.06 & 0.08\\
res2b & 436.74 & 1.12 & 108.24 & res2a\_branch2b, 3x3 & 231.20 & 0.19 & 22.78 & 0.19 & 0.19\\
res2c & 436.74 & 1.12 & 108.07 & res2a\_branch2c, 1x1 & 102.80 & 0.18 & 20.15 & 0.22 & 0.26\\
res3a & 590.88 & 1.39 & 133.09 & res2a\_branch1, 1x1 & 102.80 & 0.21 & 23.66 & 0.43 & 0.46\\
res3b & 436.74 & 1.03 & 87.72 & res3a\_branch2a, 1x1 & 51.40 & 0.09 & 7.19 & 0.09 & 0.13\\
res3c & 436.74 & 1.05 & 87.87 & res3a\_branch2b, 3x3 & 231.20 & 0.21 & 24.63 & 0.21 & 0.21\\
res3d & 436.74 & 1.04 & 88.66 & res3a\_branch2c, 1x1 & 102.80 & 0.15 & 15.18 & 0.21 & 0.25\\
res4a & 590.88 & 1.20 & 104.66 & res3a\_branch1, 1x1 & 205.50 & 0.29 & 30.35 & 0.33 & 0.39\\
res4b & 436.74 & 1.03 & 71.33 & res4a\_branch2a, 1x1 & 51.40 & 0.08 & 7.10 & 0.12 & 0.13\\
res4c & 436.74 & 1.03 & 72.52 & res4a\_branch2b, 3x3 & 231.20 & 0.20 & 23.12 & 0.21 & 0.23\\
res4d & 436.74 & 1.02 & 72.01 & res4a\_branch2c, 1x1 & 102.80 & 0.15 & 13.01 & 0.29 & 0.38\\
res4e & 436.74 & 1.02 & 72.39 & res4a\_branch1, 1x1 & 205.50 & 0.28 & 29.23 & 0.43 & 0.50\\
res4f & 436.74 & 1.02 & 71.76 & res5a\_branch2a, 1x1 & 51.40 & 0.14 & 7.61 & 0.12 & 0.19\\
res5a & 590.88 & 1.73 & 95.61 & res5a\_branch2b, 3x3 & 231.20 & 0.31 & 24.90 & 0.33 & 0.49\\
res5b & 436.74 & 1.24 & 61.55 & res5a\_branch2c, 1x1 & 102.80 & 0.27 & 12.53 & 0.47 & 0.60\\
res5c & 436.74 & 1.23 & 61.58 & res5a\_branch1, 1x1 & 205.50 & 0.51 & 30.92 & 0.52 & 0.69\\
\midrule
\textbf{Min} &  & 1.02 & 61.55 & \textbf{Min} &  & 0.06 & 5.05 & 0.06 & 0.08\\
\textbf{Max} &  & 1.73 & 133.09 & \textbf{Max} &  & 0.51 & 30.92 & 0.52 & 0.69\\
\textbf{Var} &  & 0.04 & 454.94 & \textbf{Var} &  & 0.01 & 79.42 & 0.02 & 0.03\\
\bottomrule
\end{tabular}
}
\end{table}}

\begin{figure}[h]
\centering
\includegraphics[width=\linewidth]{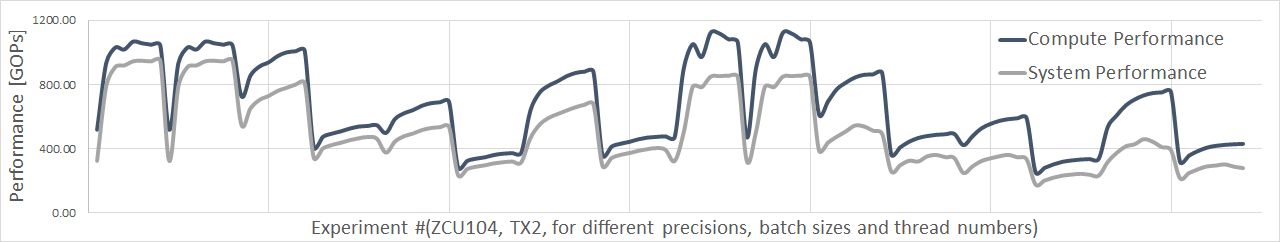}
\caption{System versus Compute Performance}
\label{fig:syscmp}
  \vspace*{-3ex}
\end{figure}

\end{document}